\documentclass[journal]{IEEEtran}
\usepackage{bbm}
\usepackage{amsfonts}
\usepackage{mathrsfs}
\usepackage{graphicx}
\usepackage{amssymb}
\usepackage{amsmath}
\usepackage{stfloats}
\usepackage{cases}
\usepackage{setspace}
\usepackage{amsthm}

\pdfoutput=1

\newtheorem{theorem}{Theorem}
\newtheorem{remark}{Remark}
\newtheorem{lemma}{Lemma}

\newtheorem{proposition}{Proposition}
\theoremstyle{plain}

\begin{document}


\title{Energy Efficiency of Cross-Tier Base Station Cooperation in Heterogeneous Cellular Networks}

\author{Weili~Nie\IEEEauthorrefmark{1}, Fu-Chun Zheng\IEEEauthorrefmark{1}, Xiaoming Wang\IEEEauthorrefmark{1}, Shi Jin\IEEEauthorrefmark{1}, Wenyi Zhang\IEEEauthorrefmark{2}\\

\thanks{\IEEEauthorrefmark{1}The authors are with the National Mobile Communications Research Laboratory, Southeast University, Nanjing 210093, China (e-mail:
\{nieweili, fzheng, wangxiaoming, jinshi\}@seu.edu.cn).}

\thanks{\IEEEauthorrefmark{2}Wenyi Zhang is with the Department of Electronic Engineering and Information Science, University of Science and Technology of China, Hefei 230027, China (e-mail:
wenyizha@ustc.edu.cn).}

\thanks{The research has been supported by the National Basic Research Program of China (973 Program) under grant 2012CB316004.}


%
%

    }

\maketitle

\begin{abstract}
Heterogeneous cellular networks (HetNets) are to be deployed for future wireless communication to meet the ever-increasing mobile traffic demand. However, the dense and random deployment of small cells and their uncoordinated operation raise important concerns about energy efficiency. Base station (BS) cooperation is set to play a key role in managing interference in the HetNets.
In this paper, we consider BS cooperation in the downlink HetNets where BSs from different tiers within the respective cooperative clusters jointly transmit the same data to a typical user, and further optimize the energy efficiency performance. First, based on the proposed clustering model, we derive the spatial average rate using tools from stochastic geometry. Furthermore, we formulate a power minimization problem with a minimum spatial average rate constraint and derive an approximate result of the optimal received signal strength (RSS) thresholds. Building upon these results, we effectively address the problem of how to design appropriate RSS thresholds, taking into account the trade-off between spatial average rate and energy efficiency. Simulations show that our proposed clustering model is more energy-saving than the geometric clustering model, and under our proposed clustering model, deploying a two-tier HetNet is significantly more energy-saving compared to a macro-only network.

\end{abstract}


\section{Introduction}

It has been reported that information and communication technology (ICT) already contributes around 2\% of the global carbon dioxide emission and this is expected to increase rapidly in the future \cite{1}. In addition to the environmental impact, the ICT infrastructure is responsible for about 10\% of the world's electric energy consumption, to which the wireless telecommunication industry is the major contributor. Therefore, an energy-efficient cellular network operation is needed more than ever before to reduce both the operational costs and carbon footprint of this industry. Recently, designing green cellular networks has received great attention amongst network operators, regulatory bodies such as 3GPP and ITU and green communications research projects such as EARTH and GreenTouch \cite{2}-\cite{5}.

Generally, macro base stations (BSs) are not designed for providing high data rates but large coverage ranges. Therefore, due to the explosive growth of mobile data traffic, an increasing portion of the mobile data and voice traffic is expected to be offloaded from the macrocell network onto other low power and low cost small cell networks \cite{SC}, resulting in heterogeneous cellular networks (HetNets) \cite{6}. HetNets comprise a conventional cellular network overlaid with a diverse set of lower-power BSs or access points (APs), such as picocells \cite{7}, femtocells \cite{8}, WiFi APs \cite{9} and perhaps relays \cite{10}. Heterogeneity is expected to be a key feature of future cellular networks, and an essential means for providing higher end-user throughput as well as expanding its indoor and cell edge coverage. Nevertheless, deployment of a large number of small cells overlaying the macrocells is not without new technical challenges \cite{eICIC}.

BS cooperation, described varyingly as coordinated multi-point (CoMP) \cite{11}, network multiple-input multiple-output (MIMO) \cite{12} or more recently, as a cloud radio access network (C-RAN) \cite{13} serves as one of effective techniques to manage inter-cell interference and improve spectral efficiency. Specifically, cooperation schemes may range from coordinated scheduling and beamforming (CS/CB) \cite{14}, \cite{15} to full joint signal processing \cite{16}, depending on the employed backhaul architecture, tolerable mobility and complexity, and other constraints. Despite falling short of their initial hype \cite{Jeffrey limits}, BS cooperation transmission is nonetheless beneficial and could require a redefinition of the different nodes in the HetNets \cite{Five disrupt}. This requires new tools for performance prediction and analysis.


\subsection{Related Work and Motivations}

Recently, a new general model for wireless node distribution based on stochastic geometry has been proposed \cite{19}-\cite{21} and the authors have shown to be a tractable and reasonably accurate solution for analyzing important metrics such as signal-to-interference-plus-noise ratio (SINR) coverage and spatial average rate.

Based on the stochastic geometry framework, a number of works have discussed BS cooperation techniques. \cite{24} investigated outage probability of BS cooperation by large-deviation theory. The BSs are clustered using a regular lattice, whereby BSs in the same cluster mitigate mutual interference by zero-forcing beamforming. \cite{25} analyzed intra-cluster interference coordination for randomly deployed BSs, considering a random clustering process where cluster stations are located according to a random point process and groups of BSs associated with the same cluster coordinate. A model for pair-wise BS cooperation in the downlink with irregular BS deployment was treated in \cite{26}. Coherent joint-transmission with power-splitting with and without additional dirty paper coding was considered.
\cite{27} presented a general model for analyzing non-coherent joint-transmission BS cooperation,
and characterized SINR distribution under user-centric clustering and channel dependent scheduling. \cite{28} studied the ergodic capacity of a multicell distributed antenna system (DAS), where remote antenna units are spread within each cell to cooperatively transmit to user terminals. \cite{NCJT} extended the work in \cite{27} towards the multi-tier HetNets and derived coverage probability.

However, little work has analyzed or optimized energy savings of BS cooperation in the HetNets (especially under different path loss exponents).
In this work, we model the location of BSs, both the macrocells and small cells, as independent Poisson point processes (PPPs) with different intensities. Based on the stochastic geometry framework, we aim at deriving a tractable result for spatial average rate. From the perspective of energy efficiency, we mainly focus on two issues: the optimal RSS thresholds of different tiers and the performance gains of our proposed clustering model in the HetNets.

\subsection{Contributions and Organizations}

\emph{A modified user-centric clustering model:}
We propose a modified user-centric clustering model, based on tier-specific received signal strength (RSS) thresholds. The main difference between our proposed clustering model and the geometric clustering model in \cite{27} and \cite{NCJT} is that we consider the impact of fading coefficient on RSS thresholds, and thus the cooperative region of each tier is a randomly shaped region.
Compared to the geometric clustering model, where the cooperative region of each tier is a deterministic two-dimensional ball, it is shown that our proposed clustering model is more energy-saving. More importantly, we assume different path loss exponents for different tiers, a fact which has not been considered in most literature.

\emph{Characterization of spatial average rate:} We analytically calculate the spatial average rate for a typical user located at the center of a cooperative cluster in a $K$-tier HetNet. Based on the stochastic geometry framework, the expression is reasonably tractable and enjoys a high degree of generality.
In addition, energy efficiency can be obtained from dividing the spatial average rate by the average intra-cluster power consumption. Building upon these results, we  effectively address the problem of how to design appropriate RSS thresholds, taking into account the trade-off between spatial average rate and energy efficiency.

\emph{Optimal RSS thresholds of tiers:} we formulate a power minimization problem under a minimum spatial average rate constraint and derive an approximated result. Simulation results demonstrate the tightness of the approximation. Moreover, the approximate result has a closed form in a few special cases.
The optimal RSS thresholds are influenced by multiple system parameters, such as deployment density, average power consumption per BS, backhaul power overhead and path loss exponents. Based on the approximate results, simulations show that the extra deployment of small cells is considerably more energy-saving compared to traditional macro-only networks.

In the remainder of this paper, Section II presents the system model. The spatial average rate expression is derived in Section III. In Section IV, we turn to optimizing the RSS thresholds for the power minimization problem. Numerical results and discussions are provided in Section V. Finally, Section VI concludes the paper.

\section{System Model}


We consider a HetNet composed by $K$  independent network tiers. For notational ease, we denote ${\cal K} = \left\{ {1,2, \cdots ,K} \right\}$. BSs across tiers differ in terms of deployment density ${\lambda _k}$, transmit power ${p_k}$, and path loss exponent $\alpha_k$ (${\alpha_k>2}$). 
The BS locations of each tier are assumed to be samples from an independent homogeneous PPP \{${{\rm{\Phi }}_k}$\}. Users are randomly distributed as a homogeneous PPP ${{\rm{\Phi }}_U}$ with intensity ${\lambda _U}$. BSs and users are equipped with a single antenna. A case of a 3-tier HetNet utilizing a mix of macro, micro and pico BSs is illustrated in Fig. 1, where each user can be served by BS cooperation from different tiers.

\begin{figure}
    \centering
    \includegraphics[width=7.6cm, height=4.0cm]{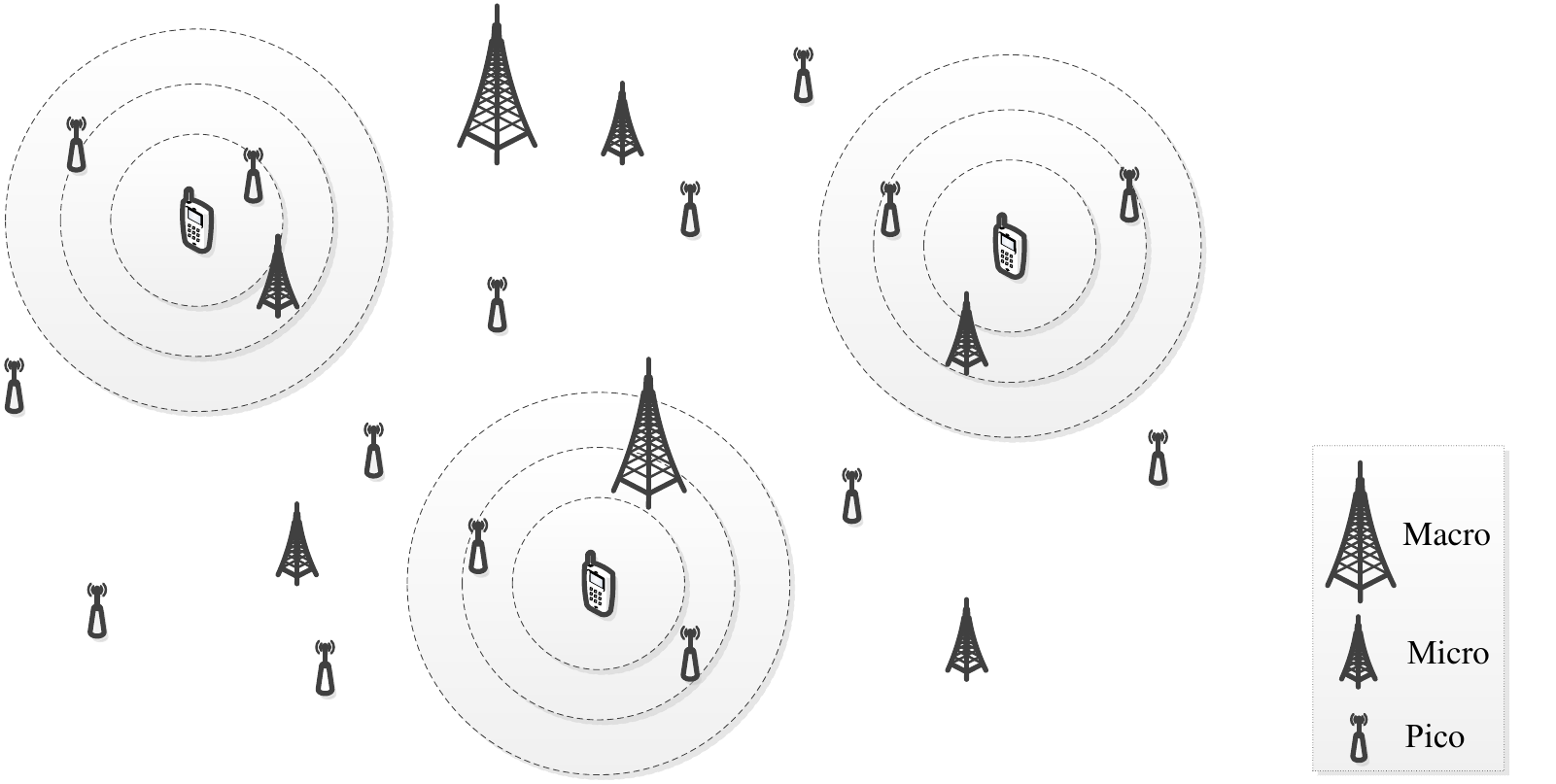}
    \caption{Illustration of a 3-tier HetNet utilizing a mix of macro, micro and pico BSs. The considered user is located in the center of each cluster, where the innermost circle denotes pico tier cooperative region, the middle circle denotes micro tier cooperative region and the outermost circle denotes macro tier cooperative region.}
\end{figure}

\subsection{Cooperation Strategy}

Without loss of generality, we focus on the downlink analysis at a typical user located at the origin $o \in {\mathbb{R}^2}$.
For this, we assume that a subset of the total ensemble of BSs cooperate by jointly transmitting the same data to this tagged receiver. This kind of BS cooperation strategy is feasible in the lightly-loaded scenario, i.e. ${\lambda _U} < \sum\nolimits_{k \in {\cal K}} {{\lambda _k}}$. As we know, network densification, driven by the rise of HetNets, brings about an interesting and novel phenomenon: each user competes only with a relatively small number of other users for a BS's service and it may even have one or more BSs to solely serve itself \cite{Five disrupt}, \cite{Jeffrey Seven}. In such a sense, the "lightly loaded scenario" assumption will be very relevant in future HetNets.

Denote by ${\cal C} \subset  \cup _{k = 1}^K{{\rm{\Phi }}_k}$ the set of the cooperative BSs for the typical user. Consequently, the link from the cooperative BSs to the typical user is a multiple-input single-output (MISO) channel.
Assuming soft combining and treating interference as noise, the SINR at the typical user is given by
\begin{align}\label{1}
{\rm{SINR}} = \frac{{\sum\limits_{x \in {\cal C}} {{p_{\nu \left( x \right)}}{\Psi _x}{{\left\| x \right\|}^{ - {\alpha _{\nu \left( x \right)}}}}} }}{{\sum\limits_{x \in \bar {\cal C}} {{p_{\nu \left( x \right)}}{\Psi _x}{{\left\| x \right\|}^{ - {\alpha _{\nu \left( x \right)}}}} + {\sigma ^2}} }},
\end{align}
where $\nu \left( x \right)$ returns the index of network tier to which the BS located at $x \in {\mathbb{R}^2}$ belongs, i.e. $\nu \left( x \right) = k$ iff $x \in {{\rm{\Phi }}_k}$, ${p_{\nu \left( x \right)}}$ is the transmit power of the BS located at $x$, $\Psi_{x}$ denotes the random fading coefficient (including shadow fading and fast fading) from the BS located at $x$ to the typical user, $\bar {\cal C} \buildrel \Delta \over =  \cup _{k = 1}^K{{\rm{\Phi }}_k}\backslash {\cal C}$ denotes the BSs that are not in the cooperative set, and ${\sigma ^2}$ is the white noise power.

\subsection{User-centric Clustering Model}

The clustering method adopted in this work is user-centric, in which the user chooses its surrounding BSs dynamically to form its serving subset. Considering the typical user located at the origin, we group BSs with sufficiently high RSS to the typical user into a cooperative cluster. Namely, the $i^{th}$ BS from the ${k^{th}}$ tier located at $x_{k,i}$ belongs to the cooperative cluster of the typical user only if ${p_k}{\Psi _{{x_{k,i}}}}{\left\| {{x_{k,i}}} \right\|^{ - {\alpha _k}}} \geqslant {T_k}$, where ${T_k}$ is the ${k^{th}}$ tier RSS threshold. In the following, we will substitute ${\Psi _{{x_{k,i}}}}$ by $\Psi _{k,i}$ for simplicity. We assume that the fading coefficients \{$\Psi_{k,i}$\} are i.i.d. random variables with the same distribution as ${{\Psi _k}}$, $k \in \mathcal{K}$. Furthermore, it is assumed that $\mathbb{E}\left[ {\Psi _k^{\frac{2}{{{\alpha _k}}}}} \right] < \infty ,{\text{  }}\forall k \in \mathcal{K}$. Thus, the set of cooperative BSs from the ${k^{th}}$  tier is
\begin{align}\label{CoMd}
\mathcal{C}{_k} = \left\{ {{x_{k,i}} \in {\Phi _k}\left| {{p_k}{\Psi _{k,i}}{{\left\| {{x_{k,i}}} \right\|}^{ - {\alpha _k}}} \geqslant {T_k}} \right.} \right\}.
\end{align}
Note that $\mathcal{C} = \sum\nolimits_{k \in \mathcal{K}} {\mathcal{C}{_k}} $ and the threshold ${T_k}$ serves as a tier-specific and tunable design parameter for optimizing the system performance.

Due to the existence of fading in the clustering model, the ${k^{th}}$ tier cooperative region turns out to be a randomly shaped region and the mean cooperative radius of the ${k^{th}}$ tier is given by
\begin{align}\label{Rk}
{R_k} = {\left( {\frac{{{p_k}}}{{{T_k}}}} \right)^{\frac{1}{{{\alpha _k}}}}}\mathbb{E}\left[ {\Psi _k^{\frac{1}{{{\alpha _k}}}}} \right].
\end{align}

\subsection{Intra-cluster Power Consumption}

In order to characterize intra-cluster power consumption, we use the linear approximation of the BS power model given by \cite{32}. Hence,
the average power consumption of a BS in the $k^{th}$ tier is given by,
\begin{align}\label{8}
    {P_{k,in}} =  {{P_{k0}} + {\Delta _k}{p _k}},
\end{align}
where ${P_{k0}}$ and ${\Delta _k}$ denote the static power expenditure and the slope of load-dependent power consumption of a BS in the $k^{th}$ tier, respectively.

Before calculating the average intra-cluster power consumption, we need to derive the average number of BSs within the ${k^{th}}$ tier cooperative set ${{\cal C}_k}$, which is obtained from the following lemma.
\begin{lemma}
 Let ${{\cal N}_k}$ be the number of BSs within the ${k^{th}}$ tier cooperative set ${{\cal C}_k}$. Then, its mean value is given by  ${N_k} = \pi {\lambda _k}{\left( {\frac{{{p _k}}}{{{T_k}}}} \right)^{\frac{2}{{{\alpha _k}}}}}\mathbb{E}\left[ {\Psi _k^{\frac{2}{{{\alpha _k}}}}} \right]$.
\end{lemma}
\emph{Proof:} See Appendix A. \hfill $\blacksquare$

Note that the cooperative BSs within a cluster have to share data for transmission, so the intra-cluster power consumption should also take backhaul overhead into account. Here we denote by ${P_{bh}}$ the backhaul power consumption per BS, and then the average intra-cluster power consumption is calculated as follows,
\begin{align}\label{10}
\begin{split}
  {P_{cl}} & = \sum\limits_{k = 1}^K {{N_k}\left( {{P_{k,in}} + {P_{bh}}} \right)}  \\
   & = \sum\limits_{k = 1}^K {\pi {\lambda _k}{{\left( {\frac{{{p _k}}}{{{T_k}}}} \right)}^{\frac{2}{{{\alpha _k}}}}}\mathbb{E}\left[ {\Psi _k^{\frac{2}{{{\alpha _k}}}}} \right]\left( {{P_{k0}} + {\Delta _k}{p_k} + {P_{bh}}} \right)}.  \\
\end{split}
\end{align}


\section{Spatial Average Rate}

%
In this section, we derive the spatial average rate expression in the $K$-tier HetNet.
We assume that appropriate adaptive modulation/coding is used, and thus the spatial average rate of a typical user (in nats/s/Hz) is
\begin{align}\label{2}
    \tau  = {\mathbb{E}_{{\text{SINR}}}}\left[ {\ln \left( {1 + {\text{SINR}}} \right)} \right].
\end{align}
Note that the expectation is both spatial and temporal averaging since the SINR of the typical user incorporates the impact of the point processes and the fading coefficients.

By substituting (\ref{CoMd}) into (\ref{1}), the SINR at the typical user under the proposed user-centric clustering model can be obtained,
and (\ref{2}) can be accordingly written as

\begin{align}\label{3}
\begin{split}
& \tau  = \\
& {\mathbb{E}_{  {\left\{ {{\Phi _k}} \right\},\left\{ {{\Psi _k}} \right\}} }} \left[ {\ln \left( {1 + \frac{{\sum\limits_{k \in {\cal K}} {\sum\limits_{{x_{k,i}} \in {{\cal C}_k}} {{p_k}{\Psi _{k,i}}{\left\| x_{k,i} \right\|}^{ - {\alpha _k}}} } }}{{\sum\limits_{k \in {\cal K}} {\sum\limits_{{x_{k,j}} \in {{\bar {\cal C}}_k}} {{p_k}{\Psi _{k,j}}{\left\| x_{k,j} \right\|}^{ - {\alpha _k}}} }  + {\sigma ^2}}}} \right)} \right],
\end{split}
\end{align}
where ${\bar {\cal C}_k} \buildrel \Delta \over = {\Phi _k}\backslash {{\cal C}_k}$ denotes the subset of non-cooperative BSs in the ${k^{th}}$ tier. The expectation is taken over both the PPP of $\left\{ {{\Phi _k}} \right\}$ and fading coefficients \{${\Psi_k}$\}.
To guarantee quality of service, we assume that the user scheduling is conducted so that the cooperative regions of the communicating users do not overlap.

Note that in (\ref{3}), we have implicitly assumed an open-access network where a user is allowed to connect to any BS in the HetNet without restriction. Other possible strategies are closed or hybrid access in which a user is allowed to connect to a subset ${\cal B} \subseteq {\cal K}$ of the tiers \cite{YiZhong}. The SINR expression for closed or hybrid access could be obtained with a similar approach to that for (\ref{3}), and the following analytical results for open access should also be modified accordingly.


In order to evaluate (\ref{3}), for completeness, we first replicate a useful lemma as follows,
\begin{lemma} [{\cite [Lemma 1] {GeCS}}]
For a homogeneous PPP ${\Phi_k} \subset {\mathbb{R}^2}$ with density ${\lambda _k}$, if each point ${x_{k,i}} \in {\Phi _k}$ is transformed to ${y_{k,i}} \in {\mathbb{R}^2}$ via ${y_{k,i}} = \Psi _{k,i}^{ - \frac{1}{{{\alpha _k}}}}{x_{k,i}}$, where $\left\{ {{\Psi _{k,i}}} \right\}$ are i.i.d. variables with $\mathbb{E}\left[ {\Psi _k^{\frac{2}{{{\alpha _k}}}}} \right] < \infty $, the new point process ${{\tilde \Phi }_k} \subset {\mathbb{R}^2}$, defined by the transformed points $\left\{ {y_{k,i}} \right\}$, is also a homogeneous PPP with density ${\tilde \lambda _k} = {\lambda _k}\mathbb{E}\left[ {\Psi _k^{\frac{2}{{{\alpha _k}}}}} \right]$.
\end{lemma}

The above lemma can be used to facilitate the inclusion of fading coefficient in (\ref{3}), by interpreting the fading effect as a random displacement of the original BS locations.

The general result of the spatial average rate is derived as follows, consisting of thermal noise as well as per-tier BS density ${\lambda _k}$, transmit power ${p_k}$, path loss exponent ${\alpha _k}$, RSS threshold $T_k$ and random fading coefficient $\Psi _{k}$ with an arbitrary distribution.
\begin{theorem}
The spatial average rate of a typical user in a $K$-tier HetNet with non-coherent joint-transmission is given by
\begin{align} \label{7}
\begin{split}
&  \tau = \int_0^\infty  {\left\{ {\exp \left[ { - \sum\limits_{k = 1}^K {\pi {\lambda _k}\mathbb{E}\left[ {\Psi _k^{\frac{2}{{{\alpha _k}}}}} \right]{{\left( {t{p_k}} \right)}^{\frac{2}{{{\alpha _k}}}}}\mathcal{Z}\left( {t,{T_k},{\alpha _k}} \right)} } \right]} \right.}  \\
&  \left. { - \exp \left[ { - \sum\limits_{k = 1}^K {\pi {\lambda _k}\mathbb{E}\left[ {\Psi _k^{\frac{2}{{{\alpha _k}}}}} \right]{{\left( {t{p_k}} \right)}^{\frac{2}{{{\alpha _k}}}}}\Gamma \left( {1 - \frac{2}{{{\alpha _k}}}} \right)} } \right]} \right\}\frac{{{e^{ - {\sigma ^2}t}}}}{t} dt,
\end{split}
\end{align}
where $\Gamma \left(  \cdot  \right)$ denotes the gamma function and
\begin{align}\label{6}
\mathcal{Z}\left( {t,{T_k},{\alpha _k}} \right) = \gamma \left( {1 - \frac{2}{{{\alpha _k}}},{T_k}t} \right) + \left( {{e^{ - {T_k}t}} - 1} \right){\left( {{T_k}t} \right)^{ - \frac{2}{{{\alpha _k}}}}}.
\end{align}
\end{theorem}

\emph{Proof:}
Introduce two auxiliary variables ${J_S}$ and ${J_I}$, which denote the received signal power from the cooperative BSs and the aggregate interference created by non-cooperative BSs, respectively, i.e.,
\begin{align}\label{25}
\begin{split}
{J_S} \buildrel \Delta \over = \sum\limits_{k \in {\cal K}} {\sum\limits_{{y_{k,i}} \in {{\cal C}_k}} {{p_k}{{\left\| {\Psi _k^{-\frac{1}{{{\alpha _k}}}}}{{x_{k,i}}} \right\|}^{ - {\alpha _k}}}} }, \\
{J_I} \buildrel \Delta \over = \sum\limits_{k \in {\cal K}} {\sum\limits_{{y_{k,j}} \in {{\bar {\cal C}}_k}} {{p_k}{{\left\| {\Psi _k^{-\frac{1}{{{\alpha _k}}}}}{{x_{k,j}}} \right\|}^{ - {\alpha _k}}}} }.
\end{split}
\end{align}

By applying Lemma 2, ${J_S}$ and ${J_I}$ can be equivalently expressed as,
\begin{align}\label{25}
\begin{split}
{J_S} = \sum\limits_{k \in {\cal K}} {\sum\limits_{{y_{k,i}} \in {{\cal D}_k}} {{p_k}{{\left\| {{y_{k,i}}} \right\|}^{ - {\alpha _k}}}} }, \\
{J_I} = \sum\limits_{k \in {\cal K}} {\sum\limits_{{y_{k,j}} \in {{\bar {\cal D}}_k}} {{p_k}{{\left\| {{y_{k,j}}} \right\|}^{ - {\alpha _k}}}} }.
\end{split}
\end{align}
where $\mathcal{D}{_k} = \left\{ {{y_{k,i}} \in {{\tilde \Phi }_k}\left| {{p_k}{{\left\| {{y_{k,i}}} \right\|}^{ - {\alpha _k}}} \geqslant {T_k}} \right.} \right\}$ and ${\bar {\cal D}_k} = {\tilde \Phi _k}\backslash {{\cal D}_k}$.
Accordingly, (\ref{3}) can be rewritten as
\begin{align}\label{NewRate}
\tau  = {{\mathbb E}_{{J_S},{J_I}}}\left[ {\ln \left( {1 + \frac{J_S}{{J_I}  + {\sigma ^2}}} \right)} \right].
\end{align}

By applying \cite [Lemma 1] {31}, we can rewrite (\ref{NewRate}) as
\begin{align}\label{26}
\begin{split}
\tau & = {{\rm E}_{{J_S},{J_I}}}\left\{ {\int_0^\infty  {\frac{{{e^{ - z}}}}{z}\left[ {1 - \exp \left( {\frac{{ - z{J_S}}}{{{J_I} + {\sigma ^2}}}} \right)} \right]dz} } \right\}\\
{\rm{ }}& \mathop = \limits^{\left( a \right)} {{\rm E}_{{J_S},{J_I}}}\left\{ {\int_0^\infty  {\frac{{{e^{ - {\sigma ^2}t}}}}{t}\exp \left( { - t{J_I}} \right)\left[ {1 - \exp \left( { - t{J_S}} \right)} \right]dt} } \right\}
\end{split}
\end{align}
where (a) follows from a change of variable $z = t\left( {{J_I} + {\sigma ^2}} \right)$. Since ${J_S}$ and ${J_I}$ are mutually independent due to ${\cal D} \cap \bar {\cal D} = \emptyset $, by applying the Fubini's theorem, we have
\begin{align}\label{FinalTau}
\tau  = \int_0^\infty  {\frac{{{e^{ - {\sigma ^2}t}}}}{t}{{\cal L}_{{J_I}}}\left( t \right)\left[ {1 - {{\cal L}_{{J_S}}}\left( t \right)} \right]dt},
\end{align}
where ${{\cal L}_{{J_S}}}\left( t \right)$ and ${{\cal L}_{{J_I}}}\left( t \right)$ are the Laplace transforms of random variables ${J_S}$ and ${J_I}$, respectively. Using the definition of the Laplace transform yields,
\begin{align}\label{27}
\begin{split}
& {{\cal L}_{{J_I}}}\left( t \right) = {\mathbb{E}_{J_I}}\left[ {\exp \left( { - tJ_I } \right)} \right] \\
& = {\mathbb{E}_{\left\{ {{{\tilde \Phi }_k}} \right\}}}\left[ {\exp \left( { - t\sum\limits_{k \in {\cal K}} {\sum\limits_{{y_{k,j}} \in {{\bar {\cal D}}_k}} {{p_k}{{\left\| {{y_{k,j}}} \right\|}^{ - {\alpha _k}}}} } } \right)} \right]\\
& \mathop  = \limits^{\left( a \right)} \prod\limits_{k = 1}^K {{\mathbb{E}_{\left\{ {{{\tilde \Phi }_k}} \right\}}}\left[ {\prod\limits_{{y_{k,j}} \in {{\bar {\cal D}}_k}} {\exp \left( { - t{p_k}{{\left\| {{y_{k,j}}} \right\|}^{ - {\alpha _k}}}} \right)} } \right]} \\
& \mathop  = \limits^{\left( b \right)} \prod\limits_{k = 1}^K {\exp \left\{ { - 2\pi {{\tilde \lambda }_k}\int_{{{\left( {\frac{{{p_k}}}{{{T_k}}}} \right)}^{\frac{1}{{{\alpha _k}}}}}}^\infty  {\left[ {1 - \exp \left( { - t{p_k}{r^{ - {\alpha _k}}}} \right)} \right]rdr} } \right\}}  \\
& \mathop  = \limits^{\left( c \right)} \exp \left[ { - \sum\limits_{k = 1}^K {\pi {\lambda _k}\mathbb{E}\left[ {\Psi _k^{\frac{2}{{{\alpha _k}}}}} \right]{{\left( {t{p_k}} \right)}^{\frac{2}{{{\alpha _k}}}}}\mathcal{Z}\left( {t,{T_k},{\alpha _k}} \right)} } \right], \\
\end{split}
\end{align}
where (a) follows from the independence property of different fading channels and BS point processes among tiers, (b) follows from the probability generating functional (PGFL) of PPP, and (c) follows from a change of variable $u = {\left( {t{p _k}} \right)^{ - 2/{\alpha _ k}}}{r^2}$ and the definition
\begin{align}\label{DefZ}
\mathcal{Z}\left( {t,{T_k},{\alpha _k}} \right) \triangleq \int_{{{\left( {{T_k}t} \right)}^{ - \frac{2}{{{\alpha _k}}}}}}^\infty  {\left[ {1 - \exp \left( { - {u^{ - \frac{{{\alpha _k}}}{2}}}} \right)} \right]du}.
\end{align}
We can calculate the above integral by making a change of variable $v = {u^{ - \frac{{{\alpha _k}}}{2}}}$, that is
\begin{align}\label{CalZ}
\begin{split}
  \mathcal{Z}\left( {t,{T_k},{\alpha _k}} \right) &= \frac{2}{{{\alpha _k}}}\int_0^{{T_k}t} {\left( {1 - {e^{ - v}}} \right){v^{ - \frac{2}{{{\alpha _k}}} - 1}}dv}  \\
 & \mathop  = \limits^{\left( a \right)} \gamma \left( {1 - \frac{2}{{{\alpha _k}}},{T_k}t} \right) + \left( {{e^{ - {T_k}t}} - 1} \right){\left( {{T_k}t} \right)^{ - \frac{2}{{{\alpha _k}}}}},
\end{split}
\end{align}
where $\gamma \left( {s,z} \right) = \int_0^z {{t^{s - 1}}{e^{ - t}}dt}$ is the lower incomplete gamma function, and (a) follows from integration by parts. Combining (\ref{27}) with (\ref{CalZ}) completes the calculation of ${{\cal L}_{{J_I}}}\left( t \right)$.
Similarly, the expression of ${{\cal L}_{{J_S}}}\left( t \right)$ is given by
\begin{align}\label{Js}
{\mathcal{L}_{{J_S}}}\left( t \right) = \exp \left[ { - \sum\limits_{k = 1}^K {\pi {\lambda _k}\mathbb{E}\left[ {\Psi _k^{\frac{2}{{{\alpha _k}}}}} \right]{{\left( {t{p_k}} \right)}^{\frac{2}{{{\alpha _k}}}}}\mathcal{Z}'\left( {t,{T_k},{\alpha _k}} \right)} } \right],
\end{align}
where $\mathcal{Z}'\left( {t,{T_k},{\alpha _k}} \right) = \Gamma \left( {1 - \frac{2}{{{\alpha _k}}}} \right) - \mathcal{Z}\left( {t,{T_k},{\alpha _k}} \right)$.

Finally, by substituting the expressions of ${{\cal L}_{{J_S}}}\left( t \right)$ and ${{\cal L}_{{J_I}}}\left( t \right)$ into (\ref{FinalTau}), we obtain the desired result in (\ref{7}).    \hfill $\blacksquare$

Although not a closed form, this expression is amenable to efficient numerical evaluation, as opposed to the usual Monte Carlo methods that rely on repeated random sampling to estimate the results.

\begin{remark}
Note that since $\mathcal{Z}\left( {t,{T_k},{\alpha _k}} \right)$ in (\ref{7}) increases with $T_k$, ${\tau}$ is a strictly monotonically decreasing function of $\left\{ {T_k} \right\}$. However, the RSS thresholds $\left\{ T_k \right\}$ cannot be arbitrarily small due to the fact that the power consumption will be unrealistically large. Thus, there exists a tradeoff between spatial average rate and energy saving.
\end{remark}

\section{Power Minimization Problem}

Based on the analytical result of spatial average rate in (\ref{7}), in this section, we want to determine the optimal RSS thresholds in the HetNet, which minimize the average intra-cluster power consumption in (\ref{10}) while satisfying the minimum spatial average rate requirement. Since the BS density is typically quite high in HetNets,
we ignore the noise and focus on the interference-limited regime.
Hence, by combining (\ref{10}) with (\ref{7}), we formulate the problem as follows,
\begin{align}\label{12}
\begin{split}
\mathop {\min } \limits_{\left\{ {{T_k}} \right\}} & \; \sum\limits_{k = 1}^K {\pi {\lambda _k}{{\left( {\frac{{{p_k}}}{{{T_k}}}} \right)}^{\frac{2}{{{\alpha _k}}}}}\mathbb{E}\left[ {\Psi _k^{\frac{2}{{{\alpha _k}}}}} \right]\left( { {{P_{k0}} + {\Delta _k}{p _k}} + {P_{bh}}} \right)} \\
{\rm{ }}\mbox{s.t.} & \;{\rm{   }} \int_0^\infty  {\frac{1}{t}\left\{ {\exp \left[ { - \sum\limits_{k = 1}^K {\pi {\lambda _k}\mathbb{E}\left[ {\Psi _k^{\frac{2}{{{\alpha _k}}}}} \right]{{\left( {t{p_k}} \right)}^{\frac{2}{{{\alpha _k}}}}}\mathcal{Z}\left( {t,{T_k},{\alpha _k}} \right)} } \right]} \right.} \\
&  \left. { - \exp \left[ { - \sum\limits_{k = 1}^K {\pi {\lambda _k}\mathbb{E}\left[ {\Psi _k^{\frac{2}{{{\alpha _k}}}}} \right]{{\left( {t{p_k}} \right)}^{\frac{2}{{{\alpha _k}}}}}\Gamma \left( {1 - \frac{2}{{{\alpha _k}}}} \right)} } \right]} \right\}dt \\
& \ge {\tau _0}
\end{split}
\end{align}
where ${\tau _0}$ denotes the minimum spatial average rate requirement.

Note that problem (\ref{12}) is an optimization problem with $K$ variables and a complicated inequality constraint.
According to \cite{34}, for a minimization problem $\mathop {\min }\limits_{{T_K},{\mathbb{T}_{ - K}}} f\left( {{T_K},{\mathbb{T}_{ - K}}} \right)$ with the vector ${\mathbb{T}_{ - K}} = \left\{ {{T_1},{T_2}, \cdots ,{T_{K - 1}}} \right\}$, if we have $\tilde f\left( {{\mathbb{T}_{ - K}}} \right) = \mathop {\min }\limits_{{T_K}} f\left( {{T_K},{\mathbb{T}_{ - K}}} \right)$, then the original problem is equivalent to $\mathop {\min }\limits_{{\mathbb{T}_{ - K}}} \tilde f\left( {{\mathbb{T}_{ - K}}} \right)$.
As a result, in the following, we divide problem (\ref{12}) into two sub-problems to solve it.

\subsection{The First Sub-problem}

When we suppose that the first $K-1$  tiers' RSS thresholds $\left\{ {{T_1},{T_2}, \cdots ,{T_{K - 1}}} \right\}$ are given, the objective function of problem (\ref{12}) is a strictly monotonically decreasing function of ${T_K}$. Thus, problem (\ref{12}) can be transformed into a sub-problem with a single variable:
\begin{align}\label{13}
\begin{split}
\mathop {{\rm{max}}}\limits_{{T_K}} & \;{\rm{  }}{T_K}\\
\mbox{s.t.} & \;{\rm{   }} \int_0^\infty  {\frac{1}{t}\left\{ {\exp \left[ { - \sum\limits_{k = 1}^K {\pi {\lambda _k}\mathbb{E}\left[ {\Psi _k^{\frac{2}{{{\alpha _k}}}}} \right]{{\left( {t{p_k}} \right)}^{\frac{2}{{{\alpha _k}}}}}\mathcal{Z}\left( {t,{T_k},{\alpha _k}} \right)} } \right]} \right.} \\
&  \left. { - \exp \left[ { - \sum\limits_{k = 1}^K {\pi {\lambda _k}\mathbb{E}\left[ {\Psi _k^{\frac{2}{{{\alpha _k}}}}} \right]{{\left( {t{p_k}} \right)}^{\frac{2}{{{\alpha _k}}}}}\Gamma \left( {1 - \frac{2}{{{\alpha _k}}}} \right)} } \right]} \right\}dt \\
& \ge {\tau _0}
\end{split}
\end{align}

Problem (\ref{13}) has a unique solution, since the left side of the constraint is a strictly monotonically decreasing function of ${T_K}$.
Instead of obtaining the optimal result numerically through the binary search algorithm, we resort to a analytically tractable lower bound, which will further be shown to be tight through numerical study in Section V.
\begin{theorem}
Given the first $K - 1$ tiers' RSS thresholds $\left\{ {{T_1},{T_2}, \cdots ,{T_{K - 1}}} \right\}$, the optimal $K^{th}$ tier RSS threshold has a lower bound $T_K^{d}$,
\begin{align}\label{14}
T_K^d = {\left( {{D_l}{e^{\frac{{{\alpha _l} - 2}}{2}C + \Theta_l  - {\tau _0}}} - \sum\limits_{k = 1}^{K-1} {{B_k}T_k^{\frac{{{\alpha _k} - 2}}{{{\alpha _k}}}}} } \right)^{\frac{{{\alpha _K}}}{{{\alpha _K} - 2}}}},
\end{align}
where $C$ is Euler's Constant, $l \in \mathcal{K}$ and


\begin{align}\label{15}
{D_l} = \frac{{\left( {{\alpha _K} - 2} \right)\Gamma \left( {1 - \frac{2}{{{\alpha _K}}}} \right)\omega _l^{\frac{{{\alpha _l}}}{2}}}}{{2{\omega _K}}},
\end{align}


\begin{align}\label{Bk}
{B_k} = \frac{{\left( {{\alpha _K} - 2} \right)\Gamma \left( {1 - \frac{2}{{{\alpha _K}}}} \right)}{\omega _k}}{{\left( {{\alpha _k} - 2} \right)\Gamma \left( {1 - \frac{2}{{{\alpha _k}}}} \right)}{\omega _K}},{\text{  }}k \in \mathcal{K},
\end{align}

\begin{align}\label{DefTheta}
\Theta_l  = \int_0^\infty  {\frac{{\exp \left( { - {\omega _l}{t^{\frac{2}{{{\alpha _l}}}}}} \right)}}{t}\left[ {1 - \exp \left( { - \sum\limits_{k = 1,k \ne l}^K {{\omega _k}{t^{\frac{2}{{{\alpha _k}}}}}} } \right)} \right]dt},
\end{align}

\begin{align}\label{wk}
{\omega _k} = \pi {\lambda _k}\mathbb{E}\left[ {\Psi _k^{\frac{2}{{{\alpha _k}}}}} \right]p_k^{\frac{2}{{{\alpha _k}}}}\Gamma \left( {1 - \frac{2}{{{\alpha _k}}}} \right),{\text{  }}k \in \mathcal{K}.
\end{align}

\end{theorem}

\emph{Proof:} See Appendix B. \hfill $\blacksquare$

Note that the selection of $l$ is arbitrary since (\ref{14}) takes the same value for any $l \in \mathcal{K}$. Besides, theorem 2 also sheds light on the optimal RSS threshold design when adding a new tier to an existing HetNet.
\begin{remark}
Fixing per-tier BS density $\lambda _k$, the lower bound of $T_K$ is an decreasing function of ${T_k}, k = 1,2, \cdots ,K-1$. It confirms our intuition that when per-tier BS density is given and the first $K - 1$ tiers' cooperative regions decrease,
the $K^{th}$ tier cooperative region should be expanded by decreasing the value of $T_K$.
\end{remark}

In addition, the lower bound of $T_K$ is also an decreasing function of $\tau_0$, yielding the following upper bound of the minimum spatial average rate.
\begin{proposition}
Given the first $K - 1$ tiers' RSS thresholds $\left\{ {{T_1},{T_2}, \cdots ,{T_{K - 1}}} \right\}$, the minimum spatial average rate $\tau_0$ has a maximum value $\tau_0^{max}$ that satisfies
\begin{align}\label{TauM}
\tau _0^{\max } = \frac{{{\alpha _l} - 2}}{2}C + {\Theta _l} - \ln \frac{{\sum\nolimits_{k = 1}^{K - 1} {{B_k}T_k^{\frac{{{\alpha _k} - 2}}{{{\alpha _k}}}}} }}{{{D_l}}}.
\end{align}
\end{proposition}

\emph{Proof:} From (\ref{14}), when ${D_l}{e^{\frac{{{\alpha _l} - 2}}{2}C + {\Theta _l} - {\tau _0}}} - {\sum\nolimits_{k = 1}^{K - 1} {{B_k}T_k^{\frac{{{\alpha _k} - 2}}{{{\alpha _k}}}}} }  \geqslant 0$, a lower bound $T_K^d$ exists. Solving this inequality gives the upper bound of $\tau_0$.   \hfill $\blacksquare$

The above proposition indicates that when the first tiers' ${K - 1}$ cooperative regions are fixed, even if the $K^{th}$ tier RSS threshold tends towards zero (i.e., the $K^{th}$ tier cooperative region expands unboundedly), a spatial average rate $\tau _0  >  \tau _0^{\max }$ cannot be achieved.

In the following, we consider two special cases where we can obtain closed-form (approximate) results of $\Theta_l$ and ${T_K^d}$.

\subsubsection{Special Case I (Two-tier HetNet)}

If we consider a two-tier HetNet, i.e., $K=2$, $\Theta_l$ can be rewritten as
\begin{align}\label{De2}
{\Theta_l} = \int_0^\infty  {\frac{{\exp \left( { - {\omega _l}{t^{\frac{2}{{{\alpha _l}}}}}} \right)}}{t}\left[ {1 - \exp \left( { - {\omega _j}{t^{\frac{2}{{{\alpha _j}}}}}} \right)} \right]} dt,
\end{align}
where $j \ne l$ and $j,l \in \left\{ {1,2} \right\}$.

Then we can transform (\ref{De2}) into a series with infinite number of terms, by applying the Taylor series expansion, as follows,
\begin{align}\label{Taylor}
\begin{split}
  \Theta_l  &= \int_0^\infty  {\frac{{\exp \left( { - {\omega _l}{t^{\frac{2}{{{\alpha _l}}}}}} \right)}}{t}\left[ {1 - \sum\limits_{n = 0}^\infty  {\frac{{{{\left( { - 1} \right)}^n}\omega _j^n{t^{\frac{{2n}}{{{\alpha _j}}}}}}}{{n!}}} } \right]} dt \\
&   = \int_0^\infty  {\exp \left( { - {\omega _l}{t^{\frac{2}{{{\alpha _l}}}}}} \right)\sum\limits_{n = 1}^\infty  {\frac{{{{\left( { - 1} \right)}^{n + 1}}\omega _j^n}}{{n!}}{t^{\frac{{2n}}{{{\alpha _j}}} - 1}}} } dt \\
 & \mathop  = \limits^{(a)} \sum\limits_{n = 1}^\infty  {\frac{{{{\left( { - 1} \right)}^{n + 1}}\omega _j^n}}{{n!}}\int_0^\infty  {{t^{\frac{{2n}}{{{\alpha _j}}} - 1}}\exp \left( { - {\omega _l}{t^{\frac{2}{{{\alpha _l}}}}}} \right)dt} }  \\
&  \mathop  = \limits^{(b)} \frac{{{\alpha _l}}}{2}\sum\limits_{n = 1}^\infty  {\frac{{{{\left( { - 1} \right)}^{n + 1}}}}{{n!}}{{\left( {{\omega _j}\omega _l^{ - \frac{{{\alpha _l}}}{{{\alpha _j}}}}} \right)}^n}\Gamma \left( {\frac{{{\alpha _l}n}}{{{\alpha _j}}}} \right)}.  \\
\end{split}
\end{align}
where (a) follows from swapping the order of integration and summation according to Fubini's theorem, and (b) follows from a change of variable $x = {\omega _l}{t^{\frac{2}{{{\alpha _l}}}}}$.

Since the indexes $l$ and $j$ are interchangeable, we can let $l=2$ if ${\omega _1}\omega _2^{ - \frac{{{\alpha _2}}}{{{\alpha _1}}}} < 1$, and $l=1$ otherwise. Without loss of generality, we assume that the system parameters satisfy ${\omega _1}\omega _2^{ - \frac{{{\alpha _2}}}{{{\alpha _1}}}} < 1$. Consequently, we only need to take the first $M$ terms to approximate the value of $\Theta_2$,
\begin{align}\label{Appr}
{\Theta_2} \approx \frac{{{\alpha _2}}}{2}\sum\limits_{n = 1}^M {\frac{{{{\left( { - 1} \right)}^{n + 1}}}}{{n!}}{{\left( {{\omega _1}\omega _2^{ - \frac{{{\alpha _2}}}{{{\alpha _1}}}}} \right)}^n}\Gamma \left( {\frac{{{\alpha _2}n}}{{{\alpha _1}}}} \right)},
\end{align}
By substituting (\ref{Appr}) into (\ref{14}), we can get a closed-form approximate result $T_2^a$, which satisfies the following expression:
\begin{align}\label{Tk2}
T_2^a = {\left( {{D_2}{e^{\frac{{{\alpha _2} - 2}}{2}C + {\Theta _2} - {\tau _0}}} - {B_1}T_1^{\frac{{{\alpha _1} - 2}}{{{\alpha _1}}}}} \right)^{\frac{{{\alpha _2}}}{{{\alpha _2} - 2}}}}.
\end{align}
The accuracy of this approximate result will be verified in Section V.

\subsubsection{Special Case II (Equal Path Loss Exponent)}

When the path loss exponents of all the tiers are identical, i.e., $\left\{ {{\alpha _k}} \right\} = \alpha $, $\Theta_l$ can be rewritten as
\begin{align}\label{DeEq}
\begin{split}
  \Theta_l & = \int_0^\infty  {\frac{{\exp \left( { - {\omega _l}{t^{\frac{2}{\alpha }}}} \right)}}{t}\left[ {1 - \exp \left( { - {t^{\frac{2}{\alpha }}}\sum\limits_{k = 1,j \ne l}^K {{\omega _k}} } \right)} \right]} dt \\
&  \mathop  = \limits^{(a)} \frac{\alpha }{2}\ln \left( {\frac{1}{{{\omega _l}}}\sum\limits_{k = 1}^K {{\omega _k}} } \right), \\
\end{split}
\end{align}
where (a) follows from \cite [Lemma 1] {31} and $\omega_k$ is defined in (\ref{wk}) with ${\alpha _k} = \alpha $.

By substituting (\ref{DeEq}) into (\ref{14}), we have
\begin{align}\label{TkEq}
T_K^d = {\left( {\Xi {e^{\frac{{\alpha  - 2}}{2}C - {\tau _0}}} - \sum\limits_{k = 1}^{K - 1} {\frac{{{\omega _k}T_k^{\frac{{\alpha  - 2}}{\alpha }}}}{{{\omega _K}}}} } \right)^{\frac{\alpha }{{\alpha  - 2}}}},
\end{align}
where
\begin{align}\label{Xi}
\Xi  = \frac{{\left( {\alpha  - 2} \right)\Gamma \left( {1 - \frac{2}{\alpha }} \right){{\left( {\sum\nolimits_{k = 1}^K {{\omega _k}} } \right)}^{\frac{\alpha }{2}}}}}{{2{\omega _K}}}.
\end{align}

From the closed-form result in (\ref{TkEq}), we can obtain an sight in the impact of the $K^{th}$ tier deployment intensity on the $K^{th}$ tier RSS threshold, which is given in the following proposition.
\begin{proposition}
In the case of equal path loss exponent, given the first $K - 1$ tiers' RSS thresholds and deployment intensities, when ${\lambda _K} \geqslant {\mathcal{G}_1}$, $T_K^d$ increases with $\lambda_K$, and when ${\lambda _K} \leqslant {\left( {{\mathcal{G}_1} - {\mathcal{G}_2}} \right)^ + }$, $T_K^d$ decreases with $\lambda_K$, where ${\left( x \right)^ + } = \left\{ {\begin{array}{*{20}{c}}
  {x,{\text{  }}x > 0} \\
  {0,{\text{  }}x \leqslant 0}
\end{array}} \right.$ and ${\mathcal{G}_1}$, ${\mathcal{G}_2}$ are given by
\begin{align} \label{M_G1}
{\mathcal{G}_1} = \frac{{2\sum\nolimits_{k = 1}^K {\mathbb{E}\left[ {\Psi _k^{\frac{2}{\alpha }}} \right]p_k^{\frac{2}{\alpha }}{\lambda _k}} }}{{\left( {\alpha  - 2} \right)\mathbb{E}\left[ {\Psi _K^{\frac{2}{\alpha }}} \right]p_K^{\frac{2}{\alpha }}}},
\end{align}
\begin{align} \label{M_G2}
{\mathcal{G}_2} = \frac{{4\sum\nolimits_{k = 1}^K {\mathbb{E}\left[ {\Psi _k^{\frac{2}{\alpha }}} \right]p_k^{\frac{2}{\alpha }}T_k^{\frac{{\alpha  - 2}}{\alpha }}{\lambda _k}} }}{{{{\left( {\alpha  - 2} \right)}^2}\Gamma \left( {1 - \frac{2}{\alpha }} \right){e^{\frac{{\alpha  - 2}}{2}C - {\tau _0}}}{{\left( {\sum\limits_{k = 1}^{K - 1} {{\omega _k}} } \right)}^{\frac{{\alpha  - 2}}{2}}}\mathbb{E}\left[ {\Psi _K^{\frac{2}{\alpha }}} \right]p_K^{\frac{2}{\alpha }}}}.
\end{align}
\end{proposition}

\emph{Proof:} See Appendix C. \hfill $\blacksquare$

Note that Proposition 2 reveals an interesting phenomenon: When the $K^{th}$ tier deployment intensity is sufficiently large, increasing $\lambda_K$ leads to a larger $K^{th}$ tier cooperative region; on the other hand, when the $K^{th}$ tier deployment intensity is sufficiently small, increasing $\lambda_K$ leads to a smaller $K^{th}$ tier cooperative region.

\subsection{The Second Sub-problem}


According to Theorem 2, by substituting (\ref{14}) into the objective function of the problem (\ref{12}) and applying some algebraic manipulations, we can formulate the second sub-problem (\ref{21}), where the constraint exists because only when ${D_l}{e^{\frac{{{\alpha _l} - 2}}{2}C + {\Theta _l} - {\tau _0}}} - {\sum\nolimits_{k = 1}^{K - 1} {{B_k}T_k^{\frac{{{\alpha _k} - 2}}{{{\alpha _k}}}}} }  \geqslant 0$ can a feasible $T_K^{d}$ exist.

\begin{align} \label{21}
\begin{split}
  \mathop {\min }\limits_{{\mathbb{T}_{ - K}}} & {\text{  }}\sum\limits_{k = 1}^{K - 1} {{\lambda _k}{B_k}\frac{{{\alpha _k} - 2}}{{{\alpha _K} - 2}}\frac{{{P_{k0}} + {\Delta _k}{p _k} + {P_{bh}}}}{{{P_{K0}} + {\Delta _K}{p _K} + {P_{bh}}}}T_k^{-\frac{2}{{{\alpha _k}}}}} \hfill \\
 &  \;\; + {\left( {{D_l}{e^{\frac{{{\alpha _l} - 2}}{2}C + {\Theta _l} - {\tau _0}}} - \sum\limits_{k = 1}^{K - 1} {{B_k}T_k^{\frac{{{\alpha _k} - 2}}{{{\alpha _k}}}}} } \right)^{\frac{2}{{2 - {\alpha _K}}}}} \hfill \\
  {\text{ }} \mbox{s.t.} & {\text{   }}{D_l}{e^{\frac{{{\alpha _l} - 2}}{2}C + {\Theta _l} - {\tau _0}}} - \sum\limits_{k = 1}^{K - 1} {{B_k}T_k^{\frac{{{\alpha _k} - 2}}{{{\alpha _k}}}}}  \geqslant 0 \hfill \\
\end{split}
\end{align}

A general result of the optimal RSS thresholds $\left\{ {{T_k^*}} \right\}$ is given as follows.
\begin{theorem}
In order to minimize the intra-cluster power consumption while satisfying the minimum spatial average rate requirement ${\tau _0}$, the optimal $K$ tiers' RSS thresholds $\left\{ {{T_k^*}} \right\}$ satisfy the following expressions:
\begin{align}\label{22}
\sum\limits_{j = 1}^K {{B_j}\Omega _{j \to k}^{\frac{{{\alpha _j} - 2}}{{{\alpha _j}}}}} {\left(T_k^*\right)}^{\frac{{{\alpha _j} - 2}}{{{\alpha _j}}}} = {D_l}{e^{\frac{{{\alpha _l} - 2}}{2}C + {\Theta _l} - {\tau _0}}},\text{ } k \in \mathcal{K},
\end{align}
where
\begin{align}\label{23}
{\Omega _{j \to k}} = \frac{{{P_{j0}} + {\Delta _j}{p_j} + {P_{bh}}}}{{{P_{k0}} + {\Delta _k}{p_k} + {P_{bh}}}}, \text{ } j,k \in \mathcal{K}.
\end{align}
Specially, when $\left\{ {{\alpha _k}} \right\} = \alpha$, a closed-form expression of $\left\{ {{T_k^*}} \right\}$ is given by
\begin{align}\label{Ropt}
{T_k^*} = {\left( {\frac{{\Xi {e^{\frac{{\alpha  - 2}}{2}C - {\tau _0}}}}}{{\sum\nolimits_{j = 1}^K {{B_j}\Omega _{j \to k}^{\frac{{\alpha  - 2}}{\alpha }}} }}} \right)^{\frac{\alpha }{{\alpha  - 2}}}}, \text{ } k \in \mathcal{K}.
\end{align}
\end{theorem}

\emph{Proof:} See Appendix D. \hfill $\blacksquare$

Note that the design of the optimal  $k^{th}$ RSS threshold jointly depends on multiple system parameters, including deployment density ${\lambda _k}$, average power consumption per BS $P_{k,in}$, backhaul power overhead $P_{bh}$ and path loss exponent $\alpha_k$. Besides, from (\ref{22}) and (\ref{Ropt}), we observe that $T_k^*$ decreases with $\tau_0$. It is intuitive that when the minimum spatial average rate becomes higher, more BSs should participate in the cooperative transmission.

Furthermore, from (\ref{Rel}) in Appendix D, the ratio of any two tiers' optimal RSS thresholds satisfies

\begin{align}\label{Rel2}
\frac{{T_j^*}}{{T_k^*}} = \frac{{{P_{j,in}} + {P_{bh}}}}{{{P_{k,in}} + {P_{bh}}}}, \text{ } j,k \in {\cal K}.
\end{align}
where $P_{k,in}$ is defined in (\ref{8}).
That is, this ratio only depends on the power parameters of these two tiers. Accordingly, we can obtain an insightful observation that when  the minimum spatial average rate $\tau_0$ increases, the most energy-efficient way to satisfy this rate requirement is to decrease each tier's RSS threshold ${T_k}$ proportionally. On the contrary, adjusting only one tier's RSS threshold is always suboptimal in terms of energy saving, even if the power consumption per BS in this tier is low.

\begin{remark}
Since (\ref{22}) is a polynomial equation, a simple binary search method or iterative method can be applied to compute the optimal result numerically. Meanwhile, due to the relationship of any two tiers' optimal RSS thresholds in (\ref{Rel2}), we just need to compute one nonlinear equation, which further reduces the complexity.
\end{remark}

By substituting the optimal RSS thresholds $\left\{ {{T_k^*}} \right\}$ into the objective function of problem (\ref{12}), the minimum intra-cluster power consumption of our proposed clustering model can be obtained.

To demonstrate that our proposed clustering model is more energy-saving than other clustering models, we consider another user-centric clustering model in \cite{NCJT} where the set of cooperative BSs from the $k^{th}$ tier is defined as ${\mathcal{C}_k} = b\left( {o,{R_k}} \right) \cap {\Phi _k}$. Here, $b\left( {o,{R_k}} \right)$ denotes a two-dimensional ball centered at origin with a radius $R_k$. We call it a geometric clustering model because the cooperative set is only determined by the Euclidean distances between BSs and users.

Before we show the minimum intra-cluster power consumption of the geometric clustering model, we need to derive the optimal cooperative radii following a similar process to that of our proposed clustering model. We omit the derivation and give the result in the following proposition.
\begin{proposition}
Assume that the fading coefficient ${\Psi _k} \sim \exp (\frac{1}{{{\mu _k}}})$. In order to minimize the intra-cluster power consumption while satisfying the minimum spatial average rate requirement ${\tau _0}$ under the geometric clustering model, the optimal $K$ tiers' cooperative radii $\left\{ {{R_k^*}} \right\}$ satisfy the following expressions:
\begin{align}\label{G1}
\sum\limits_{j = 1}^K {{{\tilde B}_j}\tilde \Omega _{j \to k}^{\frac{{{\alpha _j} - 2}}{{{\alpha _j}}}}{{\left( {R_k^*} \right)}^{\frac{{{\alpha _k}\left( {2 - {\alpha _j}} \right)}}{{{\alpha _j}}}}}}  = {\tilde D_l}{e^{\frac{{{\alpha _l} - 2}}{2}C + {\tilde \Theta _l} - {\tau _0}}},\text{ } k \in \mathcal{K},
\end{align}
where $l \in \mathcal{K}$ and
\begin{align}\label{G2}
{{\tilde B}_k} = \frac{{\left( {{\alpha _K} - 2} \right){p_k}{\mu _k}{\lambda _k}}}{{\left( {{\alpha _k} - 2} \right){p_K}{\mu _K}{\lambda _K}}},\text{ } k \in \mathcal{K},
\end{align}

\begin{align}\label{G3}
{{\tilde \Omega }_{j \to k}} = \frac{{\left( {{P_{j0}} + {\Delta _j}{p_j} + {P_{bh}}} \right){p_k}{\mu _k}}}{{\left( {{P_{k0}} + {\Delta _k}{p_k} + {P_{bh}}}, \right){p_j}{\mu _j}}},\text{ } j,k \in \mathcal{K},
\end{align}

\begin{align}\label{G4}
\tilde \Theta_l  = \int_0^\infty  {\frac{{\exp \left( { - {\tilde \omega _l}{t^{\frac{2}{{{\alpha _l}}}}}} \right)}}{t}\left[ {1 - \exp \left( { - \sum\limits_{k = 1,k \ne l}^K {{\tilde \omega _k}{t^{\frac{2}{{{\alpha _k}}}}}} } \right)} \right]dt},
\end{align}

\begin{align}\label{G5}
{{\tilde D}_l} = \frac{{\tilde \omega _l^{\frac{{{\alpha _l}}}{2}}}}{{\frac{{2\pi }}{{{\alpha _K} - 2}}{\lambda _K}{p_K}{\mu _K}}},
\end{align}

\begin{align}\label{G6}
{{\tilde \omega }_k} = \frac{{2{\pi ^2}}}{{{\alpha _k}}}\csc \left( {\frac{{2\pi }}{{{\alpha _k}}}} \right){\lambda _k}{\left( {{p_k}{\mu _k}} \right)^{\frac{2}{{{\alpha _k}}}}},\text{ } k \in \mathcal{K}.
\end{align}
\end{proposition}

By substituting the optimal cooperative radii $\left\{ {R_k^*} \right\}$ into the power consumption expression that ${{\tilde P}_{cl}} = \sum\nolimits_{k = 1}^K {\pi {\lambda _k}R_k^2\left( {{P_{k0}} + {\Delta _k}{p_k} + {P_{bh}}} \right)}$, we can obtain the minimum intra-cluster power consumption of the geometric clustering model. As a result, the comparison of our proposed clustering model and geometric clustering model is presented in next section.

\section{Numerical Results}
In this section, we present the simulation results to validate our analysis and evaluate the energy saving performance under the proposed clustering model. For clarity, we restrict our presented results to an interference-limited two-tier HetNet. For illustration, we assume that this two-tier HetNet consists of macro and pico BSs. Unless otherwise stated, the simulation parameters are presented in Table \ref{SIMULATION PARAMETERS} where the power parameters are chosen according to \cite{32}.

\begin{table}[!hbp] \footnotesize
\caption{\textsc{SIMULATION PARAMETERS}}
\label{SIMULATION PARAMETERS}       
\centering
\begin{tabular}{lcl}
\hline\noalign{\smallskip}
     Parameters &        \quad Value  \\
\noalign{\smallskip}\hline\noalign{\smallskip}
     Macro BS intensity ($\text m^{-2}$), ${\lambda _1}$&       \quad $1/\left( {{{250}^2}\pi } \right)$\\
     Pico BS intensity ($\text m^{-2}$), ${\lambda _2}$&        \quad $1/\left( {{{50}^2}\pi } \right)$\\
     Macro BS transmit powers (W), ${\mu _1}$&            \quad 20\\
     Pico BS transmit power (W), ${\mu _2}$&            \quad 0.13\\
     Macro BS static power expenditure (W), ${P_{10}}$&    \quad 130\\
     Pico BS static power expenditure (W), ${P_{20}}$&    \quad 6.8\\
     Backhaul power expenditure per BS (W), ${P_{bh}}$&    \quad 5\\
     Slope of Macro BS power consumption, ${\Delta _1}$&    \quad 4.7\\
     Slope of Pico BS power consumption, ${\Delta _2}$&    \quad 4.0\\
     Macro tier path loss exponent, ${\alpha_1}$&       \quad 4.3\\
     Pico tier path loss exponent, ${\alpha_2}$&      \quad 3.8\\
     Macro tier fading coefficient, ${\Psi_1}$&       \quad exp(1)\\
     Pico tier fading coefficient, ${\Psi_2}$&       \quad exp(1)\\
    \hline
\end{tabular}
\end{table}

In the Monte Carlo simulations, we choose a large spatial window, which is a square of 10km $\times$ 10km, and generate two independent PPPs with their respective intensities. For every realization, the fading coefficient ${\Psi _{k.i}}$ for each selected BS is independently generated according to exp(1), i.e., exponentially distributed random variable with unit mean, and the instantaneous ergodic rate is obtained via (\ref{3}). The final simulation results are obtained by averaging 10000 independent realizations.

\subsection{Validation of Theorem 1}

\begin{figure}
    \centering
    \includegraphics[width=\linewidth]{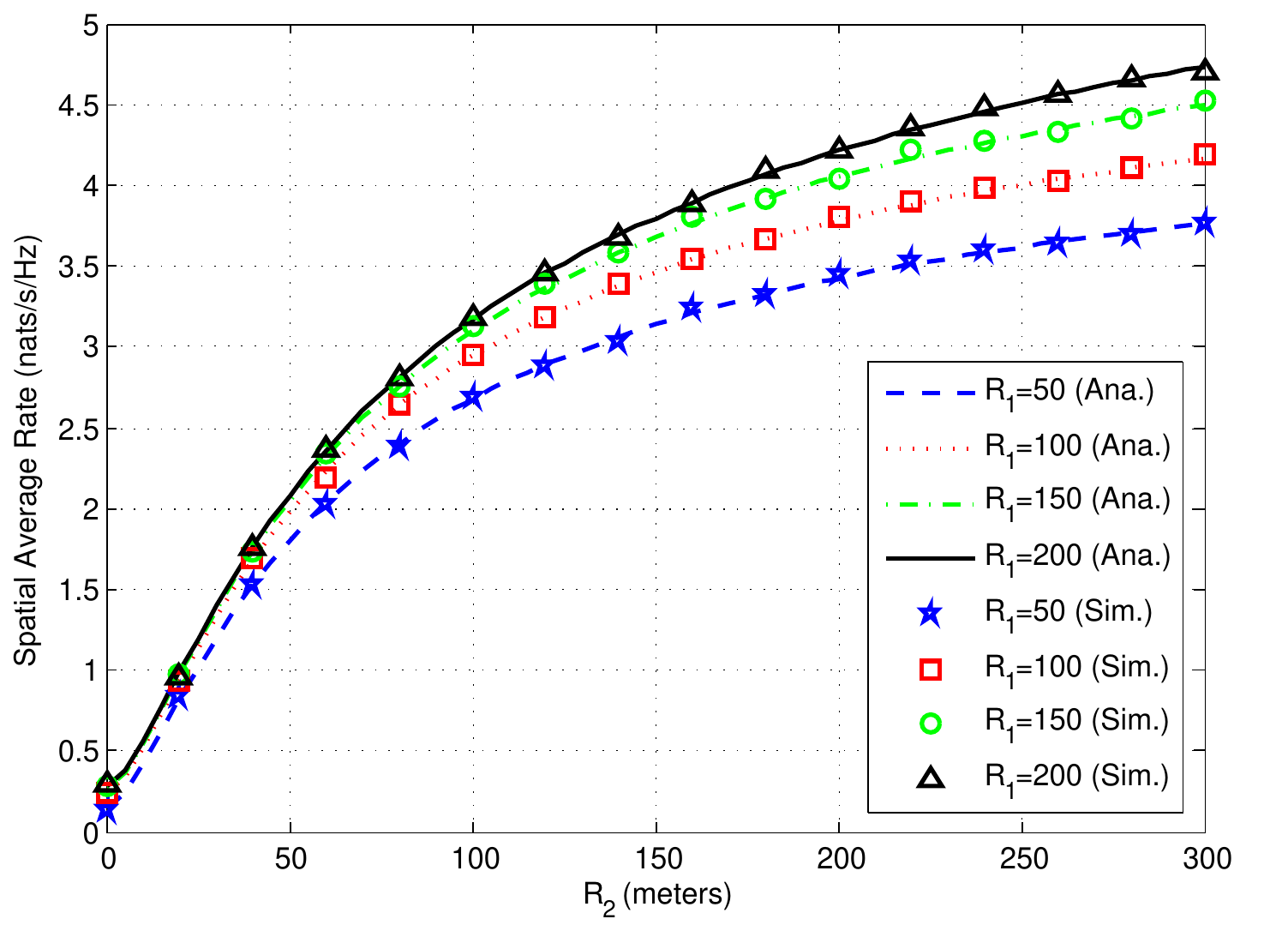}
    \caption{Spatial average rate as a function of mean cooperative radii, where ${R_1}$ and ${R_2}$ denote macro tier and pico tier mean cooperative radius, respectively.}\label{Fig3}
\end{figure}

Fig. \ref{Fig3} shows the spatial average rate of a typical user as a function of macro tier and pico tier mean cooperative radii. Note that since the mean cooperative radii $\left\{ {{R_k}} \right\}$ and RSS thresholds $\left\{ {{T_k}} \right\}$ are interchangeable according to (\ref{Rk}), here we use $\left\{ {{R_k}} \right\}$ as tunable parameters instead of $\left\{ {{T_k}} \right\}$ for a better understanding. Compared with simulation results, the analytical integrations, i.e. Eq. (\ref{7}) can be computed more efficiently and its accuracy is verified by simulations. It can be seen that the spatial average rate increases with both tiers' mean cooperative radii ${R_1}$ and ${R_2}$, since larger cooperative regions lead to higher useful signal strength and lower interference power. Besides, when the pico tier mean cooperative radius ${R_2}$ becomes sufficiently large, the increase of spatial average rate becomes more marginal. This is because the average signal strength from the pico BSs in a long distance, in general, is weak, and the impact of those distant pico BSs on spatial average rate can be neglected.

\begin{figure}
    \centering
    \includegraphics[width=\linewidth]{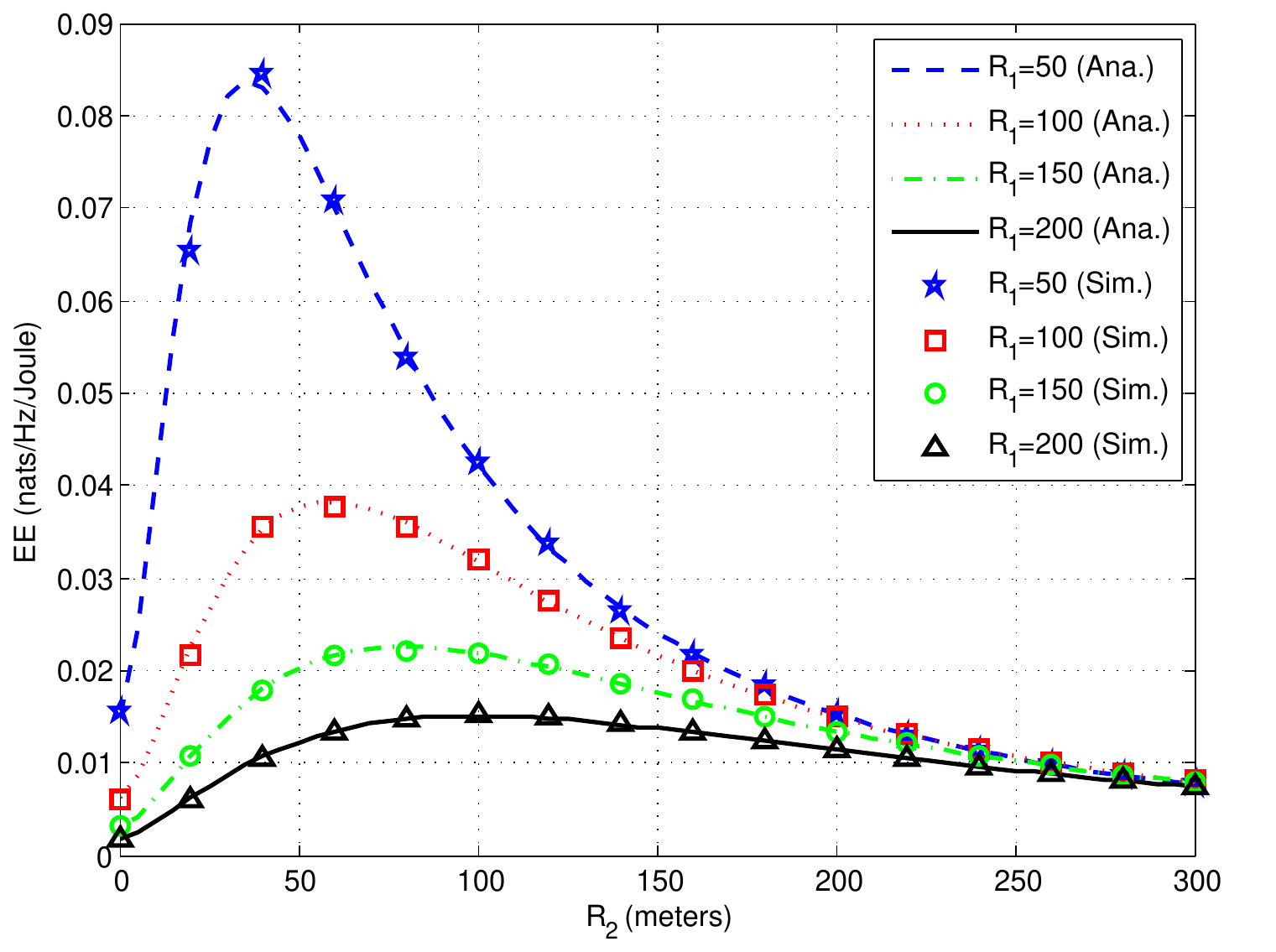}
    \caption{Energy efficiency as a function of mean cooperative radii, where ${R_1}$ and ${R_2}$ denote macro tier and pico tier mean cooperative radius, respectively.}\label{Fig4}
\end{figure}

To elucidate the trade-off between the power consumption and spatial average rate, we define the energy efficiency as ${EE} = \frac{\tau }{{{P_{cl}}}}$ (see (\ref{10}) and (\ref{7}) for the expression of $EE$). Fig. \ref{Fig4} shows the effect of macro tier and pico tier mean cooperative radii on energy efficiency. It can be seen that for a fixed ${R_1}$, there exists an optimal value $R_2^*$ which maximizes the energy efficiency; meanwhile, this optimal value $R_2^*$ increases with ${R_1}$. This result gives us an insight in designing appropriate mean cooperative radii or RSS thresholds from the perspective of energy efficiency directly.

Furthermore, we can see that energy efficiency increases as the macro tier mean cooperative radius decreases. This can be explained as follows: Compared with the positive impact on energy efficiency that more macro BSs will improve spatial average rate, the negative impact on energy efficiency that more macro BSs will increase power consumption becomes more significant.
In this figure, for example, $R_1=50, R_2=40$ is the best choice in terms of energy efficiency, but the spatial average rate for the case of $R_1=50, R_2=40$ will be very small and may not satisfy the ever-increasing rate requirement. Therefore, we should consider the tradeoff between spatial average rate and energy efficiency in a practical scenario.

\subsection{Tightness of Theorem 2}

Note that we have developed the lower bound of the optimal $K^{th}$ tier RSS threshold by fixing the first $K - 1$ tiers' RSS thresholds in Theorem 2. It is important to see how tight the lower bound is. In the following, the lower bound is calculated directly according to Theorem 2 where $\Theta_2$ ($l=2$) is calculated via (\ref{DefTheta}). The optimal value is obtained from a binary search algorithm for (\ref{7}). Also, since this is the case for $K=2$ (i.e., a two-tier HetNet) and the parameter settings satisfy that ${\omega _1}\omega _2^{ - \frac{{{\alpha _2}}}{{{\alpha _1}}}} < 1$, a closed-form approximate value can be given by (\ref{Tk2}) where $\Theta_2$  is calculated according to (\ref{Appr}) with $M=2$.

\begin{figure}
    \centering
    \includegraphics[width=\linewidth]{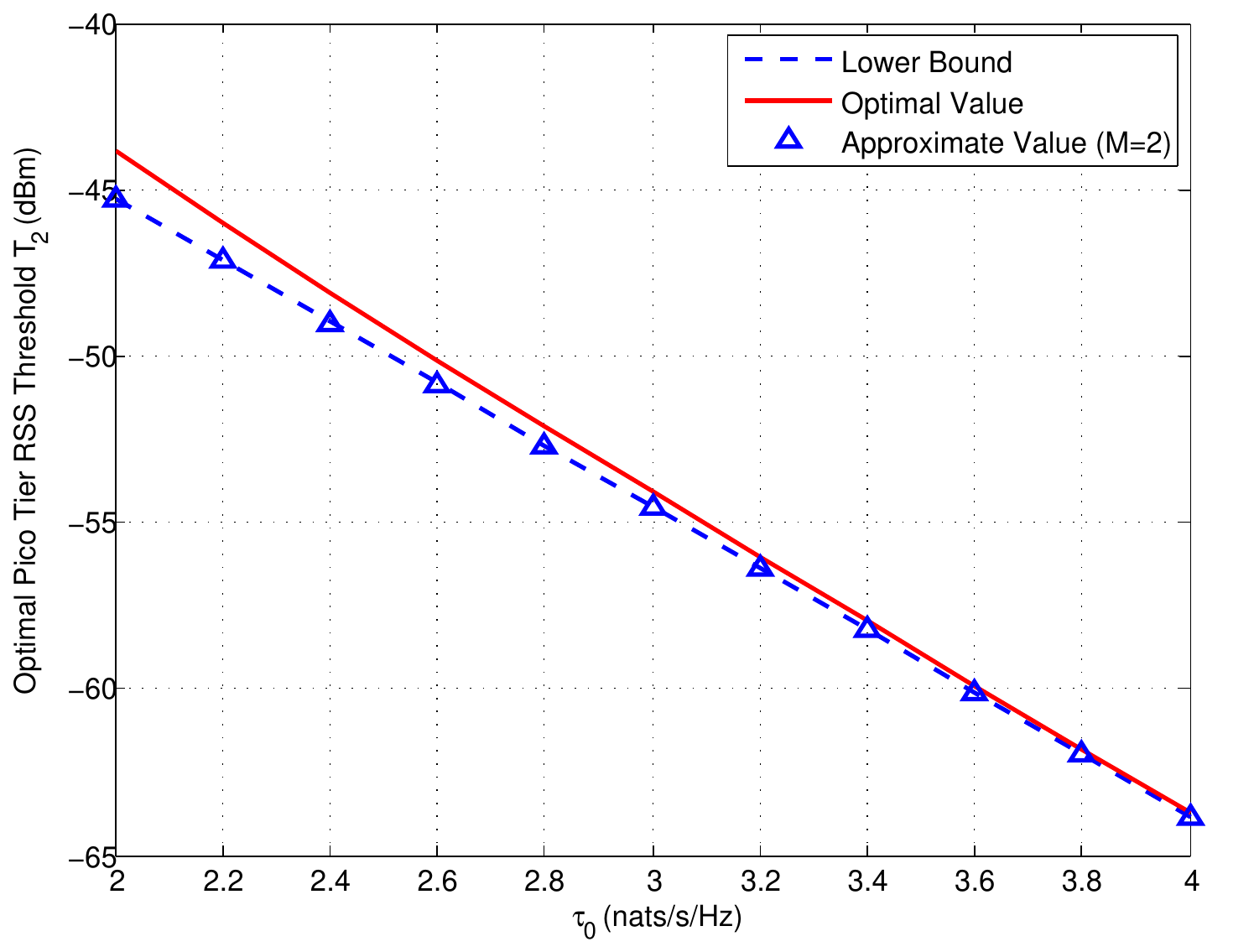}
    \caption{The optimal pico tier RSS threshold as a function of minimum spatial average rate ${\tau _0}$ where ${R_1} = 500$.}\label{Fig5}
\end{figure}

Fig. \ref{Fig5} shows the optimal pico tier RSS threshold as a function of minimum spatial average rate, given a fixed value of the macro tier mean cooperative radius, i.e. ${R_1} = 500$. First, it can be seen that the lower bound is close to the optimal value and their gap decreases as ${\tau _0}$ increases. Since the value of ${\tau _0}$ becomes increasingly large with the ever-increasing high traffic demand, this gap will be acceptable. Furthermore, we see that the approximate value is as accurate as the lower bound, which verifies the effectiveness of our approximation in the case of a two-tier HetNet. Namely, in a two-tier HetNet, we can effectively substitute the lower bound by the closed-form approximate value without losing accuracy.

\begin{figure}
    \centering
    \includegraphics[width=\linewidth]{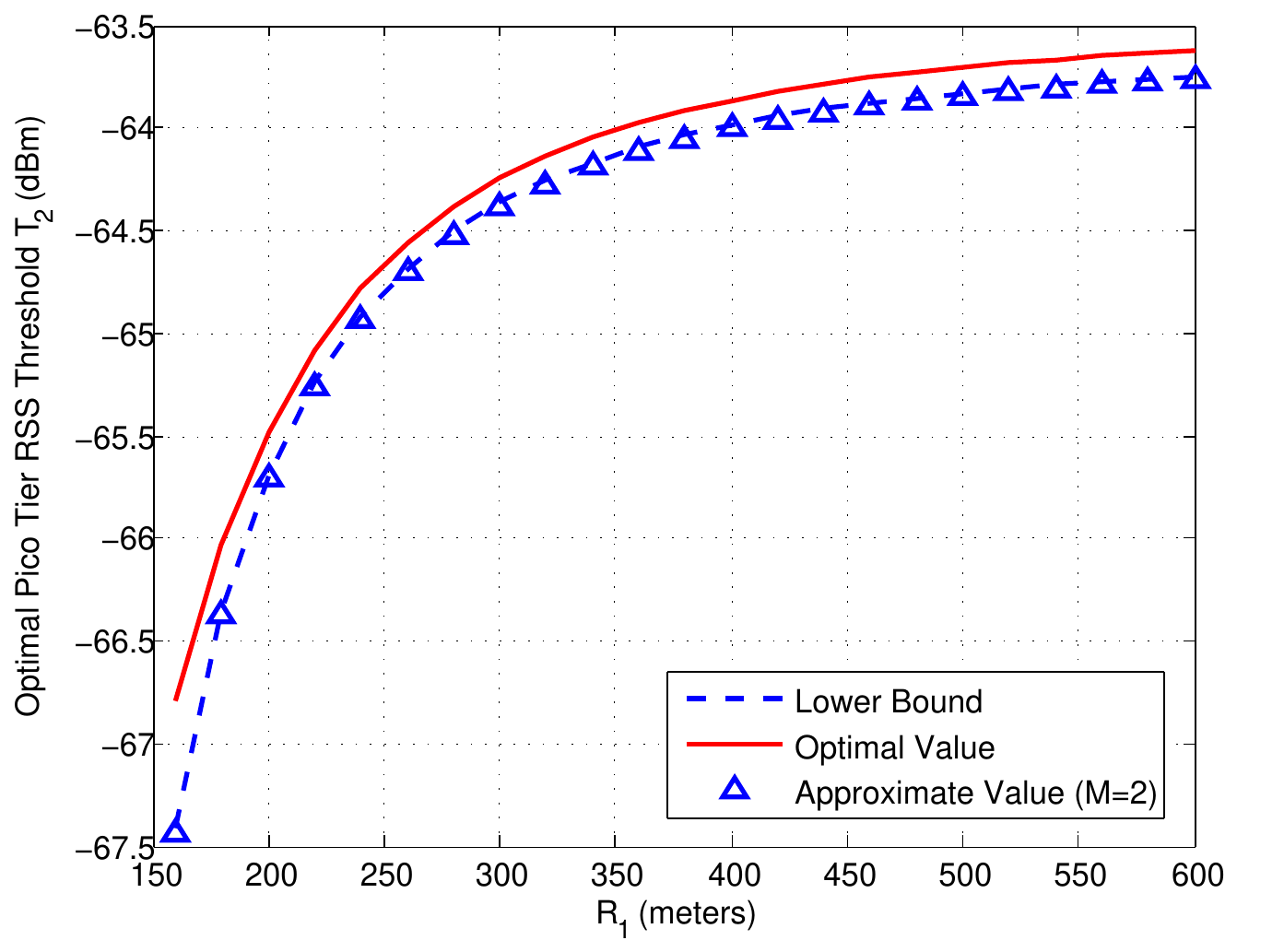}
    \caption{The optimal pico tier RSS threshold as a function of the macro tier mean cooperative radius ${R_1}$ where ${\tau _0} = 4$.}\label{Fig6}
\end{figure}

Fig. \ref{Fig6} shows the optimal pico tier RSS threshold as a function of the macro tier mean cooperative radius, given a fixed value of minimum spatial average rate, i.e. ${\tau _0} = 4$. We can see that the gap between the lower bound and the optimal value is very small. When we have ${R_1} = 300$, for instance, the gap is about 0.13 dB. More intuitively, if we consider the gap from the perspective of the optimal pico tier mean cooperative radius according to (\ref{Rk}) equivalently, the gap is about 1.3 meters, which lies within the tolerable range. Therefore, the observations above verify the effectiveness of the lower bound.  Besides, similar to Fig. \ref{Fig5}, the approximate value also matches the lower bound well, which further confirms the effectiveness of the approximation in a two-tier HetNet.

\subsection{Intra-cluster Power Consumption}

\begin{figure}
    \centering
    \includegraphics[width=\linewidth]{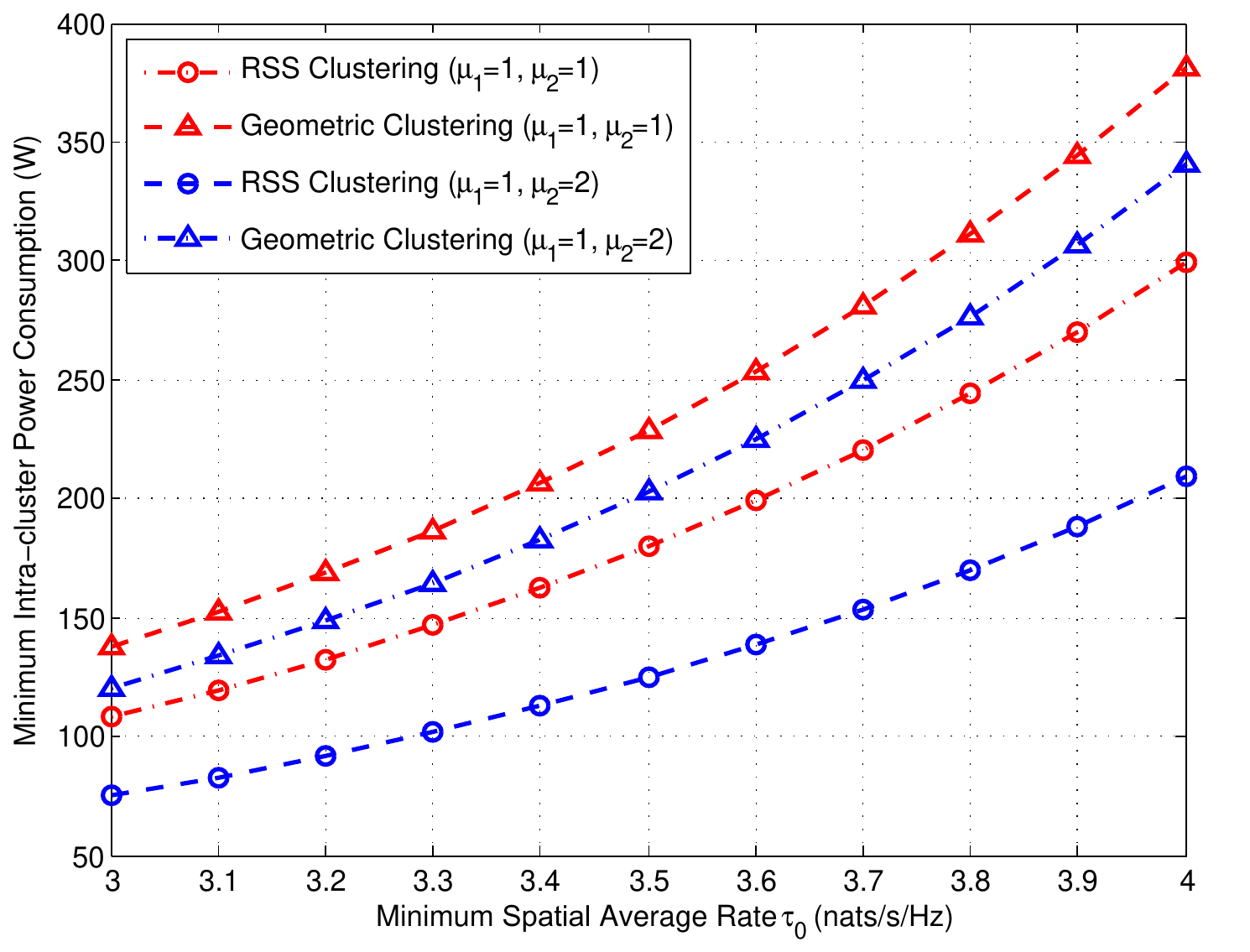}
    \caption{The minimum intra-cluster power consumption under the proposed clustering model vs. the geometric clustering model where the fading coefficient ${\Psi _k} \sim \exp (\frac{1}{{{\mu _k}}}), k=1,2$.} \label{Fig2}
\end{figure}

Fig. \ref{Fig2} shows the minimum intra-cluster power consumption varies with the minimum spatial average rate $\tau_0$, where we assume that the fading coefficient ${\Psi _k} \sim \exp (\frac{1}{{{\mu _k}}})$. The curve ``RSS Clustering'' represents our proposed clustering model, and the curve ``Geometric Clustering'' represents geometric clustering model. It can be seen that when the channel gain increases or minimum spatial average rate decreases, the minimum intra-cluster power consumption decreases, which confirms our intuition.

Furthermore, the minimum intra-cluster power consumption of our proposed clustering model is less than that of the geometric clustering model, and this energy-saving advantage of our proposed clustering model becomes more significant as the channel gain of pico tier increases. Particularly, when we set the minimum spatial average rate to be 3.5 nats/s/Hz, the minimum intra-cluster power consumption of our proposed clustering model can be reduced by about 21.5\%  and 38.3\% when $\mu_2=1$ and $\mu_2=2$, respectively. This clearly demonstrate the advantage of our proposed clustering model beyond the geometric policy. Clearly, if the channel between a nearby BS and the typical user is in a deep fading, information cannot be sent reliably over that link. Hence, this nearby BS will most probably not be considered as a serving cooperative BS. However, the geometric clustering model does not take the impact of the fading into account.

\begin{figure}
    \centering
    \includegraphics[width=\linewidth]{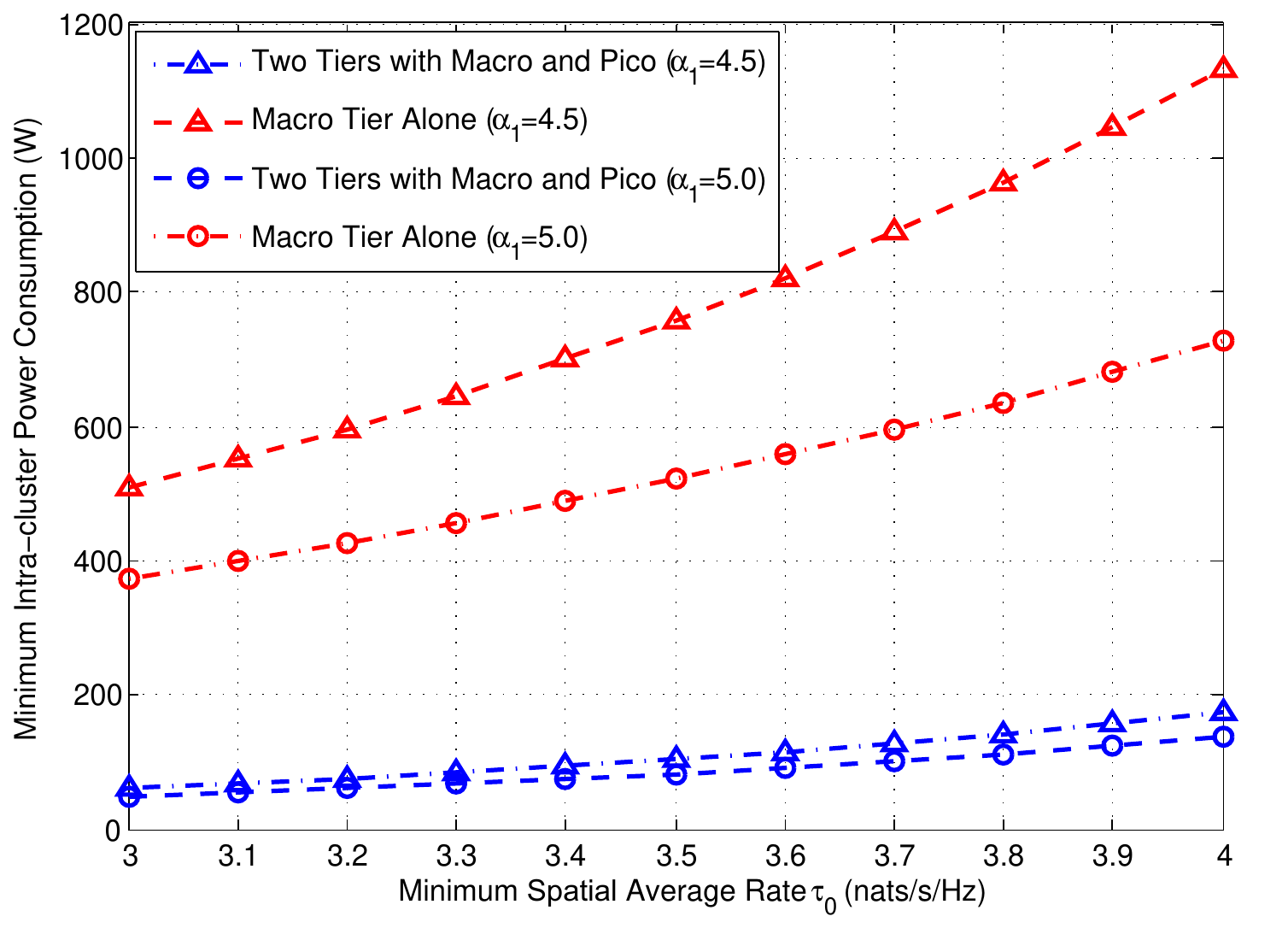}
    \caption{The minimum intra-cluster power consumption under the proposed clustering model in a two-tier HetNet vs. in a homogeneous network consisting of macro BSs alone where ${\alpha_2}= 3.8$.} \label{Fig1}
\end{figure}

Besides, Fig. \ref{Fig1} shows the minimum intra-cluster power consumption under our proposed clustering model in a two-tier HetNet vs. in a homogeneous network consisting of macro BSs alone.
We see that as the macro tier path loss exponent $\alpha_1$ increases, the intra-cluster power consumption in both cases decreases, and the influence of $\alpha_1$ in the homogeneous network is much more significant. This can be explained since the interference strength becomes smaller with the increase of macro tier path loss exponent, and thus we need fewer cooperative BSs to meet the minimum spatial average rate requirement. Furthermore, this figure shows that compared with a homogeneous network consisting of macro BSs alone, the extra deployment of pico BSs is significantly more energy-saving. Particularly, when we set the minimum spatial average rate to be 3.5 nats/s/Hz, the minimum intra-cluster power consumption of a two-tier HetNet can be reduced by about 86.3\% and 84.3\% when $\alpha_1=4.5$ and $\alpha_1=5.0$, respectively. Therefore, it verifies the effectiveness of deploying the HetNets under the proposed clustering model from the perspective of energy saving.

\section{Conclusions}
In this paper, we considered BS cooperation in the downlink HetNets where BSs from different tiers within the respective cooperative clusters jointly transmit the same data to a typical user. A user-centric clustering model, based on tier-specific RSS threshold, was proposed. We derived the spatial average rate expression for a typical user located at the center of a cooperative cluster. Furthermore, we formulated a power minimization problem with a minimum spatial average rate constraint and derived its approximate solution which was shown to be highly accurate by simulations. Building upon these results, we effectively addressed the problem of how to design appropriate RSS thresholds, taking into account the trade-off between spatial average rate and energy efficiency. Simulations showed that our proposed clustering model is more energy-saving compared to the geometric clustering model, and the extra deployment of pico BSs is significantly more energy-saving compared to the traditional macro-only network.
Note that we have not considered the detailed selection procedure and its implementation complexity involved in the user-centric clustering model, but future work on this topic is needed.

\appendix

\subsection{Proof of Lemma 1}
According to the definition of the $k^{th}$ tier cooperative set, ${{\cal N}_k}$ can be denoted as
\begin{align}\label{CalN_k}
{{\cal N}_k} = \sum\limits_{{x_{k,i}} \in {\Phi _k}} {{1_{\left\{ {{p_k}{\Psi _{k,i}}{{\left\| {{x_{k,i}}} \right\|}^{ - {\alpha _k}}} \ge {T_k}} \right\}}}},
\end{align}
where ${{\bf{1}}_{\left\{  \cdot  \right\}}}$ is an indicator function. The mean value of ${{\cal N}_k}$ can be written as
\begin{align}\label{N_k}
\begin{split}
{N_k} & ={\mathbb{E}_{{\Phi _k},{\Psi _{k,i}}}}\left[ {\sum\limits_{{x_{k,i}} \in {\Phi _k}} {{1_{\left\{ {{p_k}{\Psi _{k,i}}{{\left\| {{x_{k,i}}} \right\|}^{ - {\alpha _k}}} \ge {T_k}} \right\}}}} } \right]\\
 & \mathop  = \limits^{(a)} {\mathbb{E}_{{\Psi _{k}}}}\left[ {2\pi {\lambda _k}\int_0^\infty  {{1_{\left\{ {{p_k}{\Psi _k}{r^{ - {\alpha _k}}} \ge {T_k}} \right\}}}rdr} } \right]\\
& = \pi {\lambda _k}{\left( {\frac{{{p_k}}}{{{T_k}}}} \right)^{\frac{2}{{{\alpha _k}}}}}{\mathbb{E}}\left[ \Psi _{k}^{\frac{2}{{{\alpha _k}}}} \right],
\end{split}
\end{align}
where (a) follows from the Campbell's theorem \cite{29}.

\subsection{Proof of Theorem 2}
From (\ref{DefZ}), we have
\begin{align}\label{34}
\begin{split}
\mathcal{Z}\left( {t,{T_k},{\alpha _k}} \right) & \leqslant \int_{{{\left( {{T_k}t} \right)}^{ - \frac{2}{{{\alpha _k}}}}}}^\infty  {{u^{ - \frac{{{\alpha _k}}}{2}}}du}  \\
&  = \frac{2}{{{\alpha _k} - 2}}{\left( {{T_k}t} \right)^{\frac{{{\alpha _k} - 2}}{{{\alpha _k}}}}}, \\
\end{split}
\end{align}
where the inequality follows from the fact that $1 - {e^{ - x}} \leqslant x,\forall x \geqslant 0$. Note that the gap between two sides of this inequality diminishes when the value of $x$ becomes smaller. As a result, when $T_k$ becomes smaller, the inequality in (\ref{34}) becomes tighter.

Then by substituting (\ref{34}) into (\ref{7}) (letting ${\sigma ^2} = 0$), we can give an approximated lower bound of the spatial average rate as
\begin{align}\label{35}
\begin{split}
  \tau  & \geqslant \int_0^\infty  {\frac{1}{t}\left\{ {\exp \left[ { - t\sum\limits_{k = 1}^K {\pi {\lambda _k}{\mathbb{E}}\left[ \Psi _{k}^{\frac{2}{{{\alpha _k}}}} \right]\frac{2}{{{\alpha _k} - 2}}p_k^{\frac{2}{{{\alpha _k}}}}T_k^{\frac{{{\alpha _k} - 2}}{{{\alpha _k}}}}} } \right]} \right.}  \\
 & \left. { - \exp \left[ { - \sum\limits_{k = 1}^K {\pi {\lambda _k}{\mathbb{E}}\left[ \Psi _{k}^{\frac{2}{{{\alpha _k}}}} \right]{{\left( {t{p_k}} \right)}^{\frac{2}{{{\alpha _k}}}}}\Gamma \left( {1 - \frac{{{\alpha _k}}}{2}} \right)} } \right]} \right\}dt. \\
\end{split}
\end{align}
Before deriving the lower bound of the optimal $K^{th}$ RSS threshold, we first introduce a useful lemma \cite{33},
\begin{lemma}
Assume that $p>0,q>0$, then
\begin{align}\label{37}
\int_0^\infty  {\frac{1}{t}\left[ {\exp \left( { - {t^p}} \right) - \exp \left( { - {t^q}} \right)} \right]dt}  = \frac{{p - q}}{{pq}}C,
\end{align}
where $C$ is Euler's Constant.
\end{lemma}

By letting $\rho={\sum\limits_{k = 1}^K {\pi {\lambda _k}{\mathbb{E}}\left[ \Psi _{k}^{\frac{2}{{{\alpha _k}}}} \right]\frac{2}{{{\alpha _k} - 2}}p_k^{\frac{2}{{{\alpha _k}}}}T_k^{\frac{{{\alpha _k} - 2}}{{{\alpha _k}}}}} }$, (\ref{35}) can be simplified as
\begin{align}\label{38}
\begin{split}
  \tau  & \geqslant \int_0^\infty  {\frac{1}{t}\left[ {\exp \left( { - \rho t} \right) - \exp \left( { - \sum\limits_{k = 1}^K {{\omega _k}{t^{\frac{2}{{{\alpha _k}}}}}} } \right)} \right]} dt \\
&  \mathop  = \limits^{\left( a \right)} \int_0^\infty  {\frac{1}{t}\left[ {\exp \left( { - \rho t} \right) - \exp \left( { - {\omega _l}{t^{\frac{2}{{{\alpha _l}}}}}} \right)} \right]dt}  + \Theta_l \\
&  \mathop  = \limits^{\left( b \right)} \int_0^\infty  {\frac{1}{y}\left[ {\exp \left( { - y} \right) - \exp \left( { - {\omega _l}{\rho ^{ - \frac{2}{{{\alpha _l}}}}}{y^{\frac{2}{{{\alpha _l}}}}}} \right)} \right]dt}  + \Theta_l \\
&   = \int_0^\infty  {\frac{1}{y}\left[ {\exp \left( { - \omega _l^{\frac{{{\alpha _l}}}{2}}{\rho ^{ - 1}}y} \right) - \exp \left( { - {\omega _l}{\rho ^{ - \frac{2}{{{\alpha _l}}}}}{y^{\frac{2}{{{\alpha _l}}}}}} \right)} \right]dy}  \\
& \;\; {\text{ }} + \int_0^\infty  {\frac{1}{y}\left[ {\exp \left( { - y} \right) - \exp \left( { - \omega _l^{\frac{{{\alpha _l}}}{2}}{\rho ^{ - 1}}y} \right)} \right]dy}  + \Theta_l  \\
&  \mathop  = \limits^{\left( c \right)} \frac{{{\alpha _l} - 2}}{2}C + \ln \left( {\omega _l^{\frac{{{\alpha _l}}}{2}}{\rho ^{ - 1}}} \right) + \Theta_l, \\
\end{split}
\end{align}
where $l \in \mathcal{K}$, (a) follows from the definition of $\Theta_l$ in (\ref{DefTheta})
and (b) follows from a change of variable $y=\rho t$, (c) follows from \cite [Lemma 1] {31} and Lemma 3.
Finally, since the constraint is a strictly monotonically decreasing function of ${T_K}$, the optimization problem evolves into solving the transformed equation that
\begin{align}\label{Eq}
\frac{{{\alpha _l} - 2}}{2}C + \ln \left( {\omega _l^{\frac{{{\alpha _l}}}{2}}{\rho ^{ - 1}}} \right) + \Theta_l  = {\tau _0}.
\end{align}
By letting  ${T_K} = T_K^{d}$, we have
\begin{align}\label{39}
\begin{split}
  {\tau _0}= & \frac{{{\alpha _l} - 2}}{2}C + \ln \omega _l^{\frac{{{\alpha _l}}}{2}} + \Theta_l  \hfill \\
 &  - \ln \left[ {\sum\limits_{k = 1}^{K-1} {\frac{{2\pi }}{{{\alpha _k} - 2}}{\lambda _k}{\mathbb{E}}\left[ \Psi _{k}^{\frac{2}{{{\alpha _k}}}} \right] p_k^{\frac{2}{{{\alpha _k}}}}T_k^{\frac{{{\alpha _k} - 2}}{{{\alpha _k}}}}} } \right. \hfill \\
&  \left. { + \frac{{2\pi }}{{{\alpha _K} - 2}}{\lambda _K}{\mathbb{E}}\left[ \Psi _{K}^{\frac{2}{{{\alpha _K}}}} \right] p_K^{\frac{2}{{{\alpha _K}}}}{{\left( {T_K^d} \right)}^{\frac{{{\alpha _K} - 2}}{{{\alpha _K}}}}}} \right]. \hfill \\
\end{split}
\end{align}
Solving (\ref{39}) gives the expression of the lower bound $T_K^{d}$.

\subsection{Proof of Proposition 2}
By letting $y = {\left( {T_K^d} \right)^{\frac{{\alpha  - 2}}{\alpha }}}$, then from (\ref{TkEq}), we have
\begin{align} \label{y}
y = \delta \frac{{{{\left( {\sum\nolimits_{k = 1}^K {{\omega _k}} } \right)}^{\frac{\alpha }{2}}}}}{{{\omega _K}}} - \sum\limits_{k = 1}^{K - 1} {\frac{{{\omega _k}T_k^{\frac{{\alpha  - 2}}{\alpha }}}}{{{\omega _K}}}},
\end{align}
where $\delta  \triangleq \left( {\frac{\alpha }{2} - 1} \right)\Gamma \left( {1 - \frac{2}{\alpha }} \right){e^{\frac{{\alpha  - 2}}{2}C - {\tau _0}}}$ and
\begin{align} \label{omega2}
{\omega _k} = \pi {\lambda _k}\mathbb{E}\left[ {\Psi _k^{\frac{2}{\alpha }}} \right]p_k^{\frac{2}{\alpha }}\Gamma \left( {1 - \frac{2}{\alpha }} \right),{\text{ }}k \in \mathcal{K}.
\end{align}
Taking the derivative with respect to $y\left( {{\omega _K}} \right)$ results in
\begin{align} \label{Diffy}
\begin{split}
  \frac{{\partial y\left( {{\omega _K}} \right)}}{{\partial {\omega _K}}} = & \frac{1}{{\omega _K^2}}\left[ {\delta {{\left( {\sum\limits_{k = 1}^K {{\omega _k}} } \right)}^{\frac{\alpha }{2} - 1}} \cdot } \right. \hfill \\
 & \left. {\left( {\frac{{\alpha  - 2}}{2}{\omega _K} - \sum\limits_{k = 1}^{K - 1} {{\omega _k}} } \right) + \sum\limits_{k = 1}^{K - 1} {{\omega _k}T_k^{\frac{{\alpha  - 2}}{\alpha }}} } \right] .
\end{split}
\end{align}
First, note that $\frac{{\partial y\left( {{\omega _K}} \right)}}{{\partial {\omega _K}}} \geqslant 0$ when
\begin{align} \label{cond1}
\frac{{\alpha  - 2}}{2}{\omega _K} - \sum\nolimits_{k = 1}^{K - 1} {{\omega _k}}  \geqslant 0.
\end{align}
Since ${T_K^d}$ is a strictly monotonically increasing function of $y$ and $\omega_K$ is an affine function of $\lambda_K$, ${T_K^d}$ increases with $\lambda_K$ when $\lambda_K$ satisfies the inequality (\ref{cond1}).
Second, when $\frac{{\alpha  - 2}}{2}{\omega _K} - \sum\nolimits_{k = 1}^{K - 1} {{\omega _k}}  < 0$, from (\ref{Diffy}), we have
\begin{align} \label{Diffy2}
\begin{split}
  \frac{{\partial y\left( {{\omega _K}} \right)}}{{\partial {\omega _K}}} \leqslant & \frac{1}{{\omega _K^2}}\left[ {\delta {{\left( {\sum\limits_{k = 1}^{K - 1} {{\omega _k}} } \right)}^{\frac{\alpha }{2} - 1}} \cdot } \right. \hfill \\
 & \left. {\left( {\frac{{\alpha  - 2}}{2}{\omega _K} - \sum\limits_{k = 1}^{K - 1} {{\omega _k}} } \right) + \sum\limits_{k = 1}^{K - 1} {{\omega _k}T_k^{\frac{{\alpha  - 2}}{\alpha }}} } \right] \hfill \\
 \triangleq & h\left( {{\omega _K}} \right).
\end{split}
\end{align}
Therefore, $\frac{{\partial y\left( {{\omega _K}} \right)}}{{\partial {\omega _K}}} \leqslant 0$ when
\begin{align} \label{cond2}
h\left( {{\omega _K}} \right) \leqslant 0.
\end{align}
Similarly, we can conclude that ${T_K^d}$ decreases with $\lambda_K$ when $\lambda_K$ satisfies the inequality (\ref{cond2}).

Finally, by substituting (\ref{omega2}) into (\ref{cond1}) and (\ref{cond2}), we can get ${\mathcal{G}_1}$ and ${\mathcal{G}_2}$, respectively.

\subsection{Proof of Theorem 3}
First, we assume that $\tilde f\left( {{\mathbb{T}_{ - K}}} \right)$ is a function defined on
\begin{align}\label{ROC}
\mathcal{E}{\text{ = }}\left\{ {{\mathbb{T}_{ - K}}\left| {{D_l}{e^{\frac{{{\alpha _l} - 2}}{2}C + {\Theta _l} - {\tau _0}}} - \sum\limits_{k = 1}^{K - 1} {{B_k}T_k^{\frac{{{\alpha _k} - 2}}{{{\alpha _k}}}}}  \geqslant 0,{T_k} \geqslant 0} \right.} \right\}
\end{align}
and satisfies that
\begin{align} \label{Fun}
\begin{split}
  \tilde f\left( {{\mathbb{T}_{ - K}}} \right) = & \sum\limits_{k = 1}^{K - 1} {{\lambda _k}{B_k}\frac{{{\alpha _k} - 2}}{{{\alpha _K} - 2}}{\Omega _{k \to K}}} T_k^{\frac{2}{{{\alpha _k}}}} \hfill \\
 &  + {\left( {{D_l}{e^{\frac{{{\alpha _l} - 2}}{2}C + {\Theta _l} - {\tau _0}}} - \sum\limits_{k = 1}^{K - 1} {{B_k}T_k^{\frac{{{\alpha _k} - 2}}{{{\alpha _k}}}}} } \right)^{\frac{2}{{2 - {\alpha _K}}}}} \hfill. \\
\end{split}
\end{align}
Thus the problem (\ref{21}) can be transformed as
\begin{align} \label{Fun}
\mathbb{T}_{ - K}^* = \arg \mathop {\min }\limits_{{\mathbb{T}_{ - K}}} \tilde f\left( {{\mathbb{T}_{ - K}}} \right),
\end{align}
where $\mathbb{T}_{ - K}^* = \left\{ {T_1^*,T_2^*, \cdots ,T_{K - 1}^*} \right\}$. By setting its derivative over $T_k$ to 0 as follows,
\begin{align} \label{Der}
\begin{split}
&  \frac{{\partial \tilde f\left( {{\mathbb{T}_{ - K}}} \right)}}{{\partial {T_k}}}  = {\lambda _k}{B_k}\frac{{{\alpha _k} - 2}}{{{\alpha _K} - 2}}{\Omega _{k \to K}}\left( { - \frac{2}{{{\alpha _k}}}} \right)T_k^{ - \frac{2}{{{\alpha _k}}} - 1} \hfill \\
&  \;\; {\text{    }} - \frac{2}{{2 - {\alpha _K}}}{\left( {{D_l}{e^{\frac{{{\alpha _l} - 2}}{2}C + {\Theta _l} - {\tau _0}}} - \sum\limits_{k = 1}^{K - 1} {{B_k}T_k^{\frac{{{\alpha _k} - 2}}{{{\alpha _k}}}}} } \right)^{\frac{{{\alpha _K}}}{{2 - {\alpha _K}}}}}  \times  \hfill \\
&  \;\;{\text{    }}\frac{{{\alpha _k} - 2}}{{{\alpha _k}}}{B_k}T_k^{ - \frac{2}{{{\alpha _k}}}} \hfill \\
&   = 0, \;\; k = 1,2, \cdots,K - 1. \hfill \\
\end{split}
\end{align}
Solving the above equation, we have
\begin{align} \label{Topt1}
T_k^* = {\Omega _{k \to K}}{\left( {{D_l}{e^{\frac{{{\alpha _l} - 2}}{2}C + {\Theta _l} - {\tau _0}}} - \sum\limits_{j = 1}^{K - 1} {{B_j}{{\left( {T_j^*} \right)}^{\frac{{{\alpha _j} - 2}}{{{\alpha _j}}}}}} } \right)^{\frac{{{\alpha _K}}}{{{\alpha _K} - 2}}}},
\end{align}
where $k = 1,2, \cdots,K - 1$. Considering the relationship of $T_k^*$ and $T_j^*$, we have
\begin{align} \label{Rel}
T_j^* = {\Omega _{j \to k}}T_k^*, \;\; k,j = 1,2, \cdots,K - 1.
\end{align}
Substituting (\ref{Rel}) into (\ref{Topt1}), the result is given as
\begin{align} \label{Topt2}
\sum\limits_{j = 1}^K {{B_j}\Omega _{j \to k}^{\frac{{{\alpha _j} - 2}}{{{\alpha _j}}}}} {\left( {T_k^*} \right)^{\frac{{{\alpha _j} - 2}}{{{\alpha _j}}}}} = {D_l}{e^{\frac{{{\alpha _l} - 2}}{2}C + {\Theta _l} - {\tau _0}}},
\end{align}
where $k = 1,2, \cdots,K - 1$.
From the result of (\ref{Topt1}), the optimal values ${{\mathbb{T}_{ - K}}}$ satisfy that
\begin{align} \label{Con}
\begin{split}
  {D_l}{e^{\frac{{{\alpha _l} - 2}}{2}C + {\Theta _l} - {\tau _0}}} - \sum\limits_{j = 1}^{K - 1} {{B_j}{{\left( {T_j^*} \right)}^{\frac{{{\alpha _j} - 2}}{{{\alpha _j}}}}}}  \hfill \\
   = {\left( {T_k^*} \right)^{\frac{{{\alpha _K} - 2}}{{{\alpha _K}}}}}{\Omega _{K \to k}} \geqslant 0 \hfill , \\
\end{split}
\end{align}
which means ${{\mathbb{T}_{ - K}^*}} \in \mathcal{E}$. According to \cite{34}, since $\tilde f\left( {{\mathbb{T}_{ - K}}} \right)$ is a convex function at its domain of definition $\mathcal{E}$, the optimal value ${\mathbb{T}_{ - K}^*}$ in (\ref{Topt2}) is the solution of problem (\ref{21}).

Furthermore, by letting $k=K$ in (\ref{Topt2}), we have
\begin{align} \label{ToptK}
\begin{split}
&  \sum\limits_{j = 1}^K {{B_j}\Omega _{j \to K}^{\frac{{{\alpha _j} - 2}}{{{\alpha _j}}}}} {\left( {T_K^*} \right)^{\frac{{{\alpha _j} - 2}}{{{\alpha _j}}}}} \hfill \\
&   = \sum\limits_{j = 1}^{K - 1} {{B_j}\Omega _{j \to K}^{\frac{{{\alpha _j} - 2}}{{{\alpha _j}}}}{{\left( {T_K^*} \right)}^{\frac{{{\alpha _j} - 2}}{{{\alpha _j}}}}}}  + {\left( {T_K^*} \right)^{\frac{{{\alpha _K} - 2}}{{{\alpha _K}}}}} \hfill \\
&  \mathop  = \limits^{(a)} {D_l}{e^{\frac{{{\alpha _l} - 2}}{2}C + {\Theta _l} - {\tau _0}}} \hfill, \\
\end{split}
\end{align}
where (a) follows from (\ref{14}) in Theorem 2, then (\ref{Topt2}) holds for all $k \in \mathcal{K}$ and thus serves as an optimal solution of the original problem (\ref{12}).


\begin{thebibliography}{99}


\bibitem{1}
G. P. Fettweis and E. Zimmermann, ``ICT energy consumption-trends and challenges,'' in
\emph{Proc. 11th International Symposium Wireless Personal Multimedia Communication (WPMC)}, Sep. 2008, pp. 1-4.

\bibitem{2}
3GPP TR 32.826, Telecommunication management; Study on Energy Savings Management (ESM), (Release 10), Mar 2010.
\emph{Available:} http://www.3gpp.org/ftp/Specs/html-info/32826.htm.

\bibitem{3}
L. M. Correia, D. Zeller, O. Blume, D. Ferling, Y. Jading, I. Godor, G. Auer, and L. van der Perre, ``Challenges and enabling technologies for energy aware mobile radio networks,''
\emph{IEEE Commun. Mag.}, vol. 48, no. 11, pp. 66-72, Nov. 2010.

\bibitem{4}
X. Wang, A. V. Vasilakos, M. Chen, Y. Liu, and T. T. Kwon, ``A survey of green mobile networks: opportunities and challenges,''
\emph{ACM/Springer J. Mobile Networks and Applications}, vol. 17, no. 1, Feb. 2012.

\bibitem{5}
Z. Hasan, H. Boostanimehr, and V. K. Bhargava, ``Green cellular networks: a survey, some research issues and challenges,''
\emph{IEEE Commun. Surveys Tutorials}, vol. 13, no. 4, pp. 524-540, Fourth Quarter, 2011.

\bibitem{SC}
T. Q. S. Quek, G. de la Roche, I. Guvenc, and M. Kountouris,
\emph{Small Cell Networks: Deployment, PHY Techniques, and Resource AllocaBon,} Cambridge University Press, 2013.

\bibitem{6}
Qualcomm, ``LTE advanced: heterogeneous networks,'' white paper, Jan. 2011.

\bibitem{7}
X. Lagrange, ``Multitier cell design,''
\emph{IEEE Commun. Mag.}, vol. 35, no. 8, pp. 60-64, Aug. 1997.

\bibitem{8}
V. Chandrasekhar, J. G. Andrews, and A. Gatherer, ``Femto networks: a survey,''
\emph{IEEE Commun. Mag.}, vol. 46, no. 9, pp. 59-67, Sep. 2008.

\bibitem{9}
Qualcomm, ``A comparison of LTE-Advanced HetNets and WiFi,'' white paper.
\emph{Available:} http://goo.gl/BFMFR, Sep. 2011.

\bibitem{10}
J. Sydir and R. Taori, ``An evolved cellular system architecture incorporating relay stations,''
\emph{IEEE Commun. Mag.}, vol. 47, no. 6, pp. 115-121, June 2009.

\bibitem{eICIC}
D. Lopez-Perez, I. Guvenc, G. de la Roche, M. Kountouris, T. Q. S. Quek, and J. Zhang, ``Enhanced intercell interference coordination challenges in heterogeneous networks,''
\emph{IEEE Wireless Commun. Mag.}, vol. 18, no. 3, pp. 22-30, June 2011.

\bibitem{11}
3GPP, ``Coordinated multi-point operation for LTE physical layer aspects,'' TR 36.819, Tech. Rep., Sep. 2011.

\bibitem{12}
D. Gesbert, S. Hanly, H. Huang, S. Shamai, O. Simeone, and W. Yu, ``Multi-cell MIMO cooperative networks: a new look at interference,''
\emph{IEEE J. Sel. Areas Commun.}, vol. 28, no. 9, pp. 1380-1408, Dec. 2010.

\bibitem{13}
``C-RAN: The road towards green RAN,''
\emph{China Mobile Res. Inst.}, Beijing, China, White Paper, ver. 2.5, Oct. 2011.

\bibitem{14}
M. K. Karakayali, G. J. Foschini, and R. A. Valenzuela, ``Network coordination for spectrally efficient communications in cellular systems,''
\emph{IEEE Wireless Commun. Mag.}, vol. 13, no. 4, pp. 56-61, Apr. 2006.

\bibitem{15}
H. Dahrouj and W. Yu, ``Coordinated beamforming for the multicell multiantenna wireless system,''
\emph{IEEE Trans. Wireless Commun.}, vol. 9, no. 5, pp. 1748-1759, May 2010.

\bibitem{16}
M. Sawahashi, Y. Kishiyama, A. Morimoto, D. Nishikawa, and M. Tanno, ``Coordinated multipoint transmission/reception techniques for LTE-advanced [coordinated and distributed MIMO],''
\emph{IEEE Wireless Commun. Mag.}, vol. 17, no. 3, pp. 26-34, June 2010.

\bibitem{Jeffrey limits}
A. Lozano, R. W. Heath, and J. G. Andrews, ``Fundamental limits of cooperation,''
\emph{IEEE Trans. Inf. Theory}, vol. 59, no. 9, pp. 5213-5226, Sep. 2013.

\bibitem{Five disrupt}
F. Boccardi, R. W. Heath, A. Lozano, T. L. Marzetta, and P. Popovski, ``Five disruptive technology directions for 5G,''
arXiv preprint arXiv:1312.0229, Dec. 2013.



\bibitem{19}
J. G. Andrews, F. Baccelli, and R. K. Ganti, ``A tractable approach to coverage and rate in cellular networks,''
\emph{IEEE Trans. Commun.}, vol. 59, no. 11, pp. 3122-3134, Nov. 2011.

\bibitem{20}
H. S. Dhillon, R. K. Ganti, F. Baccelli, and J. G. Andrews, ``Modeling and analysis of K-tier downlink heterogeneous cellular networks,''
\emph{IEEE J. Sel. Areas Commun.}, vol. 30, no. 3, pp. 550-560, Apr. 2012.

\bibitem{21}
H.-S. Jo, Y. J. Sang, P. Xia, and J. G. Andrews, ``Heterogeneous cellular networks with flexible cell association: a comprehensive downlink SINR analysis,''
\emph{IEEE Trans. Wireless Commun.}, vol. 11, no. 10, pp. 3484-3495, Oct. 2012.

%

\bibitem{24}
K. Huang and J. G. Andrews, ``An analytical framework for multicell cooperation via stochastic geometry and large deviations,''
\emph{IEEE Trans. Inf. Theory}, vol. 59, no. 4, pp. 2501-2516, Apr. 2013.

\bibitem{25}
S. Akoum and R. W. Heath, ``Interference coordination: random clustering and adaptive limited feedback,''
\emph{IEEE Trans. Signal Process.}, vol. 61, no. 7, pp. 1822-1834, Apr. 2013.

\bibitem{26}
A. Giovanidis and F. Baccelli, ``A stochastic geometry framework for analyzing pairwise-cooperative cellular networks,''
arXiv preprint arXiv:1305.6254, May 2013.

\bibitem{27}
R. Tanbourgi, S. Singh, J. G. Andrews, and F. K. Jondral, ``A tractable model for non-coherent joint-transmission base station cooperation,''
arXiv preprint arXiv:1308.0041, Aug. 2013.

\bibitem{28}
Y. Lin and W. Yu, ``Ergodic capacity analysis of downlink distributed antenna systems using stochastic geometry,''
in \emph{Proc. IEEE Intl. Conf. Commun.}, June 2013, pp. 3338-3343.

\bibitem{NCJT}
R. Tanbourgi, S. Singh, J. G. Andrews, and F. K. Jondral, ``Analysis of non-coherent joint-transmission cooperation in heterogeneous cellular networks,''
arXiv preprint arXiv:1402.2707, Feb. 2014.

\bibitem{29}
D. Stoyan, W. S. Kendall, and J. Mecke,
\emph{Stochastic Geometry and Its Applications,} 2nd edition. John Wiley and Sons, 1996.

\bibitem{Jeffrey Seven}
J. Andrews, ``Seven ways that HetNets are a cellular paradigm shift,''
\emph{IEEE Commmun. Mag.}, vol. 51, no. 3, pp. 136-144, Mar. 2013.


\bibitem{YiZhong}
Y. Zhong and W. Zhang, ``Multi-channel hybrid access femtocells: a stochastic geometric analysis,''
\emph{IEEE Trans. Commun.}, vol. 61, no. 7, pp. 3016-3026, July 2013.

\bibitem{GeCS}
H. S. Dhillon and J. G. Andrews, ``Downlink rate distribution in heterogeneous cellular networks under generalized cell selection,''
\emph{IEEE Wireless Commun. Letters}, vol. 3, no. 1, pp. 42-45, Feb. 2014.

\bibitem{31}
K. Hamdi, ``Capacity of MRC on correlated Rician fading channels,''
\emph{IEEE Trans. Commun.}, vol. 56, no. 5, pp. 708-711, May 2008.

\bibitem{32}
 G. Auer, V. Giannini, C. Desset, I. Godor, P. Skillermark, M. Olsson, M. A. Imran, D. Sabella, M. J. Gonzalez, O. Blume, and A. Fehske, ``How much energy is needed to run a wireless network?,''
\emph{IEEE Wireless Commun. Mag.}, vol. 18, no. 5, pp. 40-49, Oct. 2011.


\bibitem{33}
A. Jeffrey and D. Zwillinger,
\emph{Table of Integrals, Series, and Products.} Academic Press, 2007.

\bibitem{34}
S. Boyd and L. Vandenberghe,
\emph{Convex Optimization.} Cambridge University Press, 2004.


\end{thebibliography}


\end{document}